\documentclass[journal]{IEEEtran}
\usepackage{amsmath,amsfonts}
\usepackage{algorithmic}
\usepackage{algorithm}
\usepackage{array}
\usepackage{caption}
\usepackage{subcaption}
\usepackage{textcomp}
\usepackage{stfloats}
\usepackage{url}
\usepackage{verbatim}
\usepackage{graphicx}
\usepackage{cite}
\usepackage{color}
\begin{document}

\title{Joint Channel Estimation and Computation Offloading in Fluid Antenna-assisted MEC Networks}

\author
{\IEEEauthorblockN{Ying~Ju,~\IEEEmembership{Member,~IEEE}, Mingdong~Li, Haoyu~Wang,  Lei~Liu,~\IEEEmembership{Member,~IEEE}, Youyang~Qu, Mianxiong~Dong,~\IEEEmembership{Senior~Member,~IEEE}, Victor~C.~M.~Leung,~\IEEEmembership{Life~Fellow,~IEEE}, and Chau~Yuen,~\IEEEmembership{Fellow,~IEEE}
% ,Qingqi~Pei,~\IEEEmembership{Senior~Member,~IEEE}
}

% \thanks{Ying Ju, Mingdong Li, Lei Liu, and Qingqi Pei are with the School of Telecommunications Engineering, Xidian University, Xi'an 710071, China (e-mail: juyingtju@163.com; 23011211135@stu.xidian.edu.cn; leiliu@xidian.edu.cn; qqpei@mail.xidian.edu.cn).

%     }
\thanks{Ying Ju, Mingdong Li, and Lei Liu are with the School of Telecommunications Engineering, Xidian University, Xi'an 710071, China (e-mail: juyingtju@163.com; 23011211135@stu.xidian.edu.cn; leiliu@xidian.edu.cn).

Haoyu Wang is with the Center for Pervasive Communications and Computing, University of California at Irvine, Irvine, CA 92697, USA (e-mail: haoyuw30@uci.edu).

Youyang Qu is with the Key Laboratory of Computing Power Network and Information Security, Ministry of Education, Shandong Computer Science Center (National Supercomputer Center in Jinan), Qilu University of Technology (Shandong Academy of Sciences), Jinan 250014, China (e-mail: quyy@sdas.org).

Mianxiong Dong is with the Department of Information and Electronic Engineering, Muroran Institute of Technology, Muroran 050-8585, Japan (e-mail: mx.dong@csse.muroran-it.ac.jp).

Victor C. M. Leung is with the Artificial Intelligence Research Institute, Shenzhen MSU-BIT University, Shenzhen 518172, China, also with the College of Computer Science and Software Engineering, Shenzhen University, Shenzhen 518060, China, and also with the Department of Electrical and Computer Engineering, The University of British Columbia, Vancouver, BC~V6T~1Z4, Canada (e-mail: vleung@ieee.org).

Chau Yuen is with the School of Electrical and Electronics Engineering, Nanyang Technological University, Nanyang 639798, Singapore (e-mail: chau.yuen@ntu.edu.sg).
}
}

% The paper headers
% \markboth{Journal of \LaTeX\ Class Files,~Vol.~14, No.~8, August~2024}%
% {Shell \MakeLowercase{\textit{et al.}}: A Sample Article Using IEEEtran.cls for IEEE Journals}

% \IEEEpubid{0000--0000/00\$00.00~\copyright~2021 IEEE}
% Remember, if you use this you must call \IEEEpubidadjcol in the second
% column for its text to clear the IEEEpubid mark.

\maketitle

% Dueling Double Deep Q Network (D3QN)-twin Delayed Deep Deterministic Policy Gradient (TD3)

\begin{abstract}
With the emergence of fluid antenna (FA) in wireless communications, the capability to dynamically adjust port positions offers substantial benefits in spatial diversity and spectrum efficiency, which are particularly valuable for mobile edge computing (MEC) systems. Therefore, we propose an FA-assisted MEC offloading framework to minimize system delay. This framework faces two severe challenges, which are the complexity of channel estimation due to dynamic port configuration and the inherent non-convexity of the joint optimization problem. Firstly, we propose Information Bottleneck Metric-enhanced Channel Compressed Sensing (IBM-CCS), which advances FA channel estimation by integrating information relevance into the sensing process and capturing key features of FA channels effectively. Secondly, to address the non-convex and high-dimensional optimization problem in FA-assisted MEC systems, which includes FA port selection, beamforming, power control, and resource allocation, we propose a game theory-assisted Hierarchical Twin-Dueling Multi-agent Algorithm~(HiTDMA) based offloading scheme, where the hierarchical structure effectively decouples and coordinates the optimization tasks between the user side and the base station side. Crucially, the game theory effectively reduces the dimensionality of power control variables, allowing deep reinforcement learning (DRL) agents to achieve improved optimization efficiency. Numerical results confirm that the proposed scheme significantly reduces system delay and enhances offloading performance, outperforming benchmarks. Additionally, the IBM-CCS channel estimation demonstrates superior accuracy and robustness under varying port densities, contributing to efficient communication under imperfect CSI.
\end{abstract}

\begin{IEEEkeywords}
Fluid antenna, compressed sensing, channel estimation, mobile edge computing, computation offloading, deep reinforcement learning, multi-agent.
\end{IEEEkeywords}

\section{Introduction}
\subsection{Background}
\IEEEPARstart{W}{ith} the rapid development of communication technology, especially the deployment of 5G and the upcoming 6G network, we have entered the era of the Internet of Things~(IoT), and the access volume and data computing volume of mobile devices are in an explosive growth stage, which puts forward higher requirements for higher transmission rates and computing capabilities. To address these challenges, mobile edge computing (MEC) has become a promising solution by offloading computationally intensive tasks from users to edge servers deployed at the network edge~\cite{mec0}. MEC reduces the data load on core networks compared to traditional cloud computing, easing backhaul pressure and accelerating response for delay-sensitive tasks. It also lowers system delay, saves energy, enhances service quality, and improves cache efficiency, drawing widespread attention from researchers.

To further improve the performance of MEC systems, especially in reducing transmission delay, the fluid antenna (FA) technology has emerged as a promising solution. FA is also referred to as movable antenna~\cite{intro_ma} and flexible antenna~\cite{flexible}. FA can dynamically adjust the radiation characteristics, including gain, directivity, and frequency response, by changing the distribution or state of the radiating elements~\cite{fas}. This adaptability enables it to provide optimal performance in different environments and requirements, offering great potential to further reduce MEC transmission delay through intelligent radiation adjustment. Currently, the studies have explored FA-based port selection and beamforming strategies to improve communication efficiency. For example, the authors in~\cite{fa_lo} utilize an alternating algorithm to jointly optimize port locations and bandwidth allocation to maximize sum rates, while the authors in~\cite{DL_ma_ps} achieve multi-beamforming in multicast scenarios by utilizing a deep learning (DL) algorithm to optimize the antenna position vector (APV) and antenna weight vector (AWV) at the FA base station (BS). In~\cite{fa_mec}, the authors propose an iterative algorithm that combines the interior point method and particle swarm optimization (IPPSO) to jointly refine FA port positions, user offloading ratios, and BS CPU frequency, demonstrating the potential of FA in reducing delay for MEC systems.

% port selection in FA design is one of the primary research directions. In \cite{fa_lo}, the authors utilize an alternating algorithm to jointly optimize the port location of FA and bandwidth allocation, aiming to maximize the overall sum rate. In \cite{DL_ma_ps}, the authors achieve multi-beamforming in multicast scenarios by utilizing a deep learning (DL) algorithm to optimize the antenna position vector (APV) and antenna weight vector (AWV) at the FA base station (BS). Given the inherent advantages of FA, the authors demonstrate its potential to address key challenges in MEC, especially in reducing system delay \cite{fa_mec}. Specifically, they design an alternating iterative algorithm based on the interior point method and particle swarm optimization (IPPSO). This algorithm iteratively refines the FA position, the user offloading ratio, and the BS CPU operating frequency to minimize the total delay. However, their approach relies on traditional optimization techniques and does not optimize the beamforming and transmit power design. 
However, despite its promise, the FA-assisted MEC offloading framework faces two critical challenges. The first is accurate channel estimation for fluid antenna system (FAS), which is complicated by the high dimensionality of the antenna port space and the limited sampling budget. The second is the optimization of high-dimensional, non-convex decision problems that arise from the dynamic nature of FA port positions and the stringent real-time requirements for MEC offloading decisions.

For FA channel estimation, the main challenge lies in the high-dimensional CSI caused by the large number of ports in FAS, which must be inferred from low-dimensional effective observations, which is known as the channel subspace sampling bottleneck~\cite{hur2013millimeter}. Conventional pilot-based methods require training overhead that scales with the number of FA ports~\cite{liang2019semi}, which severely degrades spectral efficiency. To overcome this, numerous studies~\cite{maCompressedSensingBased2023,newChannelEstimationReconstruction2024,xiaoChannelEstimationMovable2023,wangEstimationChannelParameters2023,zhangSuccessiveBayesianReconstructor2023,hur2013millimeter,liang2019semi} have modeled FA channels as sparse multipath parameterized channels, using compressed sensing (CS) and related techniques to exploit angular sparsity for efficient CSI recovery. Among them, the authors in~\cite{maCompressedSensingBased2023} introduces a CS-based approach that models the channel in a sparse angular domain using field response information (FRI). It performs stepwise estimation of multipath components to extract angles of departure (AOD), arrival (AOA), and the path response matrix (PRM) from limited-location measurements. In contrast, \cite{zhangSuccessiveBayesianReconstructor2023} proposes a successive Bayesian reconstructor (S-BAR) using Gaussian process regression with experiential kernels. This method exploits spatial correlation among densely deployed ports for high-accuracy CSI reconstruction, while substantially reducing pilot overhead via kernel-driven sampling and inference.

In parallel, deep learning (DL) techniques have demonstrated strong potential for channel estimation in complex, high-dimensional systems. For instance, \cite{he2018deep} proposes the learned denoising approximate message passing (LDAMP) algorithm, which models the channel matrix as a noise-corrupted image and applies a convolutional neural network (CNN) for nonlinear denoising. Although this method outperforms traditional AMP algorithms, it retains fixed-layer parameters and introduces considerable computational complexity due to the CNN structure. To enhance efficiency, \cite{wei2020deep} develops Gaussian Mixture Learned AMP (GM-LAMP), integrating a Gaussian mixture prior into the shrinkage function. This approach achieves better performance than conventional CS-based estimators while maintaining complexity similar to AMP. However, most DL-based CS methods are tailored for massive MIMO scenarios and struggle to adapt to the dynamic spatial behavior inherent in FAS. Moreover, current CS network designs typically focus on minimizing reconstruction loss based on channel estimation error but ignore the control over the measurement process and fail to jointly optimize the sampling process. This compromises both robustness and generalization in practical FA deployment.

% To address these challenges, recent advancements in computer vision research have introduced a novel training methodology grounded in information bottleneck principles - the Information Bottleneck Metric (IBM)~\cite{leeInformationBottleneckMeasurement2022} that enable compression-aware training of sensing networks by preserving critical information in the compressed measurements. This concept offers a promising direction for designing intelligent sensing modules for FA channel estimation.

For the optimization of FA-assisted MEC offloading systems, which entails a set of highly coupled decision variables, such as port selection, beamforming, and resource allocation, that must be jointly optimized under stringent constraints. These problems are typically high-dimensional, non-convex, and involve both discrete and continuous variables, leading to prohibitive computational complexity. As a result, traditional convex optimization techniques and heuristic algorithms often fall short, particularly in dynamic, large-scale communication scenarios. This calls for more adaptive and scalable approaches to tackle the complexity of such optimization tasks effectively.

To overcome these limitations, DL has been introduced as a data-driven alternative that can approximate complex non-linear mappings without explicit mathematical models. In particular, the authors in~\cite{8395149} present an innovative methodology that integrates machine learning (ML) with coordinated beamforming to effectively cope with the challenges imposed by dynamic channel variations and user mobility, offering improved adaptability in real-time signal processing. In a similar vein, the authors in~\cite{9831405} propose a predictive framework employing a hybrid neural network structure, aiming to alleviate the issues of excessive computational delay in auxiliary task offloading scenarios, where resource availability and network conditions are subject to frequent fluctuations. However, traditional supervised DL requires large-scale labeled datasets, which are often impractical to obtain in MEC environments where optimal actions are unknown or expensive to compute. In contrast, reinforcement learning (RL) provides a model-free framework that learns optimal decision policies through direct interaction with the environment. The authors in~\cite{10175025} propose two RL-based algorithms to minimize the total time exceeding the quality baseline of each task while ensuring that all tasks meet their soft quality deadlines. Furthermore, the authors in~\cite{9916276} propose an RL-driven scheme targeting the joint optimization of resource allocation and task scheduling, thereby effectively reducing delay overhead by dynamically coordinating computing and networking resources. These classical RL algorithms are well-suited to low-dimensional Markov decision processes (MDPs), but face scalability issues in high-dimensional continuous spaces due to the curse of dimensionality and convergence instability. To address this, deep reinforcement learning (DRL) combines the representation power of DL with the decision-making capability of RL, enabling the efficient learning of control policies in large and continuous action-state spaces and offers a promising approach to address complex non-convex problems in FA-assisted MEC offloading system due to its robust data analysis and executing capability~\cite{deep}. The authors in~\cite{wangFluidAntennaSystem2024} propose for the first time a joint optimization scheme based on deep reinforcement learning to realize the joint design of port selection and precoding of a multi-user MIMO downlink integrated sensing and communication (ISAC) system in a two-dimensional multi-port fluid antenna system, aiming to maximize the total communication rate while satisfying perception constraints. The authors in~\cite{ju2023joint} propose a DRL-based joint secure offloading and resource allocation (SORA) scheme that rapidly adapts to highly dynamic MEC networks, enhancing system delay performance while increasing secrecy probability through cooperation. In addition, the authors in~\cite{huang2019deep} propose a DRL-based online offloading framework for wireless powered MEC, which learns binary offloading decisions from experience and achieves near-optimal performance with low complexity. More recently, the authors in~\cite{hu2023achieving} introduce an environment-adaptive DRL framework that incorporates size and setting adaptive schemes to enable fast adaptation in dynamic MEC networks, outperforming state-of-the-art DRL and meta-learning methods. Nevertheless, many practical MEC systems involve multiple agents (e.g., users, access points, and edge servers), each with their own objectives and partial observations. This naturally motivates the use of multi-agent deep reinforcement learning (MADRL), which extends DRL to scenarios with decentralized actors and potentially conflicting interests. In~\cite{10261304}, the authors employ a MADRL-based approach to optimize the joint behavior of agents, thereby maximizing their long-term utility and enhancing the overall communication quality. These MADRL methods are particularly beneficial in FA-assisted MEC offloading systems where multiple users must jointly optimize offloading decisions and antenna configurations in a spatially distributed and temporally dynamic setting. However, existing MADRL applications primarily target general MEC or wireless resource allocation tasks, and to date, no research has specifically explored the use of DRL techniques to address the unique challenges of FA architecture optimization.

To further enhance the efficiency of DRL, game theory~\cite{mkiramweni2019survey} provides a theoretical basis for high-dimensional decision-making optimization in complex systems by modeling the decision-making behaviors of all parties in the game process. By modeling the interactions among distributed decision-makers, such as users and BSs, game theory allows for the design of decentralized and scalable optimization strategies in the presence of conflicting objectives and limited information. In particular, power control in multi-user MEC scenarios poses a challenging problem due to mutual interference, heterogeneous channel conditions, and dynamic resource availability. Game-theoretic approaches are well-suited for such settings, as they naturally model the competitive yet interdependent nature of user decisions regarding transmission power and offloading strategies. For instance, the authors in~\cite{DUPA} utilize game theory to investigate the data, user, and power allocation (DUPA) problem by modeling the DUPA problem as a potential game in the MEC environment, demonstrating its effectiveness in maximizing the number of served users and their overall data rate.

\subsection{Motivations and Contributions}
In conclusion, the current research landscape in MEC offloading optimization reveals three critical limitations. First, conventional systems employing fixed antennas lack the flexibility to adapt to dynamically changing multi-user spatial distributions, thereby constraining system performance in complex propagation environments. Second, existing DRL-based CS methods for channel estimation lack a systematic framework for evaluating the effectiveness of sensing measurements, resulting in suboptimal accuracy in signal reconstruction. Third, traditional single-agent optimization architectures face inherent challenges in handling high-dimensional decision spaces and achieving timely coordination when multiple users and resources are involved, particularly under stringent latency and resource constraints. Although there has been increasing interest in DRL and FA, no existing work has integrated DRL to jointly optimize FA reconfiguration, channel estimation, and MEC offloading decisions within a unified framework. To fill this gap, this paper proposes a novel solution that addresses these interrelated challenges.
\begin{itemize}
    \item [1)] 
    We propose a novel offloading framework that integrates FA into the MEC system, where the BS is equipped with multiple FA (FA-BS) to serve multiple users. This architecture improves transmission performance by exploiting the inherent diversity and multiplexing gains offered by FA, thereby enabling the joint optimization of port selection, beamforming, user transmit power, and MEC resource allocation with the objective of minimizing overall system delay.
    \item [2)] 
    We propose a novel channel estimation strategy, termed Information Bottleneck Metric-enhanced Channel Compressed Sensing (IBM-CCS), which is particularly well-suited for the reconfigurable and spatially flexible characteristics of FAS. This method leverages an IBM-based~\cite{leeInformationBottleneckMeasurement2022} importance generator to quantify the relevance of features obtained from the sensing network, and incorporates image-based CS technology to accurately reconstruct high-dimensional FA channels with low pilot overhead.
    \item [3)] 
    We propose a game theory-assisted DRL scheme to minimize the system delay of the MEC offloading by jointly optimizing the port selection of the FAs, the beamforming matrix, the power design of the users, and the MEC computation resource allocation. Game theory is leveraged to transform the optimization variable of user power design into price factor selection. This reduces the dimension of optimization variables by modeling the multi-user power design as a non-cooperative game. 
    \item [4)] 
    To manage the high-dimensional, non-convex optimization challenges in the system, a Hierarchical Twin-Dueling Multi-agent Algorithm~(HiTDMA) is developed. In this hierarchical structure, Twin-critic-based Base station Agent~(TBA) handles continuous control for the first layer, while the Dueling-based User Agent~(DUA) handles discrete selection for the second layer. By integrating continuous and discrete decision-making, HiTDMA enables efficient offloading decisions. Numerical simulation results verify that the proposed scheme can significantly improve the efficiency of MEC.
\end{itemize}

\subsection{Organization of the Paper}
This paper is organized as follows. Section~II describes the system model of the FA-assisted MEC network. Section~III outlines the IBM-CCS channel estimation scheme. Section~IV introduces the details of game theory for power design. Section~V introduces the details of the multi-agent computation offloading scheme based on HiTDMA. Section~VI gives the simulation results, and Section~VII concludes the paper.

\section{System Model}
This section presents the fundamental system framework considered in this paper, including the architecture of the FA-assisted MEC offloading system, the underlying FA channel model, the computation task offloading model, and the mathematical formulation of the associated optimization problem.
\subsection{FA-assisted MEC Offloading Architecture}
\begin{figure}[htbp]
    \centering
    \includegraphics[width=0.48\textwidth]{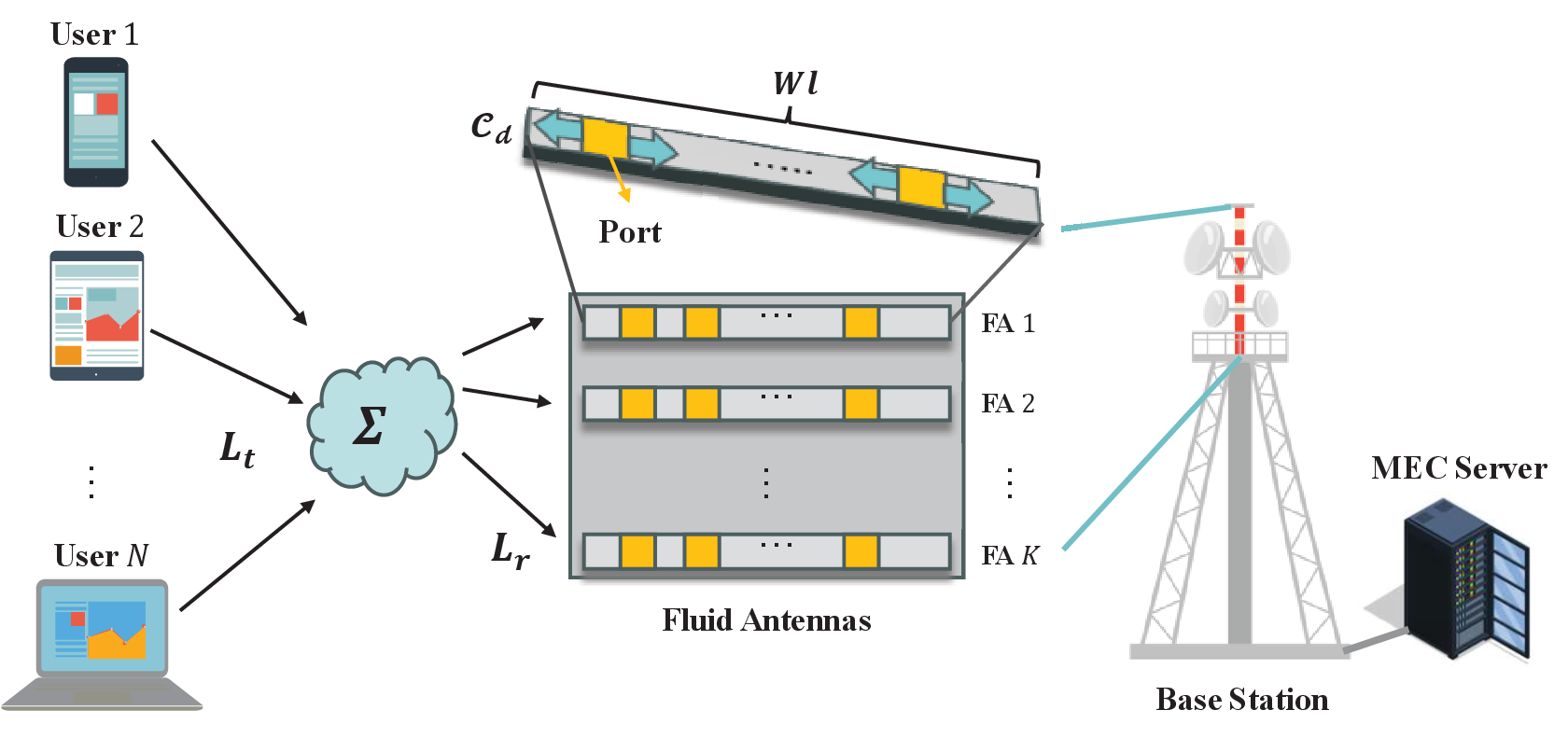} % 调整宽度
    \caption{The illustration of FA-assisted MEC network.}
    \label{fig:scene}
\end{figure}
As shown in Fig. \ref{fig:scene}, we consider an FA-assisted MEC offloading system consisting of an MEC server, $N$ users, and a BS equipped with $K$ FAs. Each user establishes a communication connection with one FA receiver of the FA-BS, and $N \le K$. The FA can switch its $N_p$ radiating elements in a linear space $\mathcal{C}_d$ of length $Wl$, where $l$ is the wavelength. The position of each radiating element can be switched to one of $M$ preset positions (or ports) uniformly distributed along the linear space. Due to the linearity of FA, the port distribution of the $n$-th user corresponding to the FA receiver can be expressed by the APV $\mathbf{d}_n =  \left [ d_1,d_2,\cdot \cdot \cdot,d_{N_p} \right ]^T \in\mathcal{C}_d$.
\par The received signal at the FA-BS can be described as
\begin{equation}
    \mathbf{y}=\mathbf{W}^H\mathbf{H}(\tilde{\mathbf{d}})\mathbf{P}^{\frac{1}{2}}\mathbf{x}+\mathbf{W}^H\mathbf{n}\label{received} 
\end{equation}
where $\mathbf{W}=\left [ \mathbf{w}_1,\mathbf{w}_2,\cdot \cdot \cdot ,\mathbf{w}_N\right ] \in \mathbb{C}^{N_p\times N}$ is the receive beamforming matrix at the FA-BS, $\mathbf{H}(\tilde{\mathbf{d}})=\left [  \mathbf{h}_1(\tilde{\mathbf{d}}),\mathbf{h}_2(\tilde{\mathbf{d}}),\cdot \cdot \cdot ,\mathbf{h}_N(\tilde{\mathbf{d}})\right ]\in \mathbb{C}^{N_p\times N}$ is the channel matrix from all $N$ users to the $N_p$ ports at the FA-BS with $\tilde{\mathbf{d}}=\left [  \mathbf{d}_1,\mathbf{d}_2,\cdot \cdot \cdot ,\mathbf{d}_N\right ] \in \mathbb{C}^{N_p\times N}$, which denoting the APV for FAs. $\mathbf{P}^{\frac{1}{2}} = \mathrm{diag} \left \{ \sqrt{p_1},\sqrt{p_2},\cdot \cdot \cdot ,\sqrt{p_N}  \right \}$ is the power matrix which representing the transmit power of users. $\mathbf{x}$ is the independent and identically distributed (i.i.d.) transmit signal vector of users, and satisfies $\mathbb{E} (\mathbf{x}\mathbf{x}^H)=\mathbf{I}_N$. $\mathbf{n }\sim \mathcal{CN} (0,\sigma ^2\mathbf{I}_{N_p})$ is the zero mean additive white Gaussian noise (AWGN) with covariance matrix $\sigma ^2\mathbf{I}_{N_p}$.
\subsection{FA Channel model}
The close spacing of FA ports enables spatial channel correlation through phase differences induced by port separation distance $\Delta d$. To address mutual coupling effects while maintaining reconfigurability, adjacent ports maintain a minimum separation of $l/2$.

% Since the mobile area of FAs at the FA-BS is much smaller than the signal propagation distance, we assume that the farfield condition is satisfied between the users and the BS. We employ a field response-based channel model, where the channel vector between the $n$-th user and the FA-BS is expressed as

Given that the motion range of FAs at the FA-BS is significantly smaller than the typical signal propagation distance, the farfield condition is assumed to hold for the user-to-BS links. A field response-based parametric channel model is adopted, wherein the small-scale fading characteristics are represented through estimated channel parameters. Specifically, the uplink channel vector between the $n$-th user and the FA-BS is modeled as
\begin{equation}
\tilde{\mathbf{h}}_n(\mathbf{d}_n)=\rho_n\tilde{\mathbf{g}}_n\odot \mathbf{\alpha }_n (\mathbf{d}_n,\theta_n ),\label{channel_equation}
\end{equation}
where $\rho_n$ denotes the large-scale fading component accounting for path loss, and $\odot$ denotes Hadamard product. $\mathbf{\alpha }_n (\mathbf{d}_n,\theta_n )=\left [ e^{j\frac{2\pi}{l}d_1\cos(\theta_n)},e^{j\frac{2\pi}{l}d_2\cos(\theta_n)},\cdots, e^{j\frac{2\pi}{l}d_{N_p}\cos(\theta_n)} \right ]^T$ represents the field-response vector of the received channel paths, dependent on the port positions $\mathbf{d}_n$ and the AoA $\theta_n$. The vector $\tilde{\mathbf{g}}_n=\left [ \tilde{g}_{n,1},\tilde{g}_{n,2},\cdots ,\tilde{g}_{n,N_p} \right ]^T$ represents the estimated small-scale fading coefficients between the $n$-th user and the FA ports. These coefficients are not directly measurable and are instead inferred through channel estimation techniques.

Based on the above model, the signal-to-interference-noise ratio (SINR) from the $n$-th user to the FA-BS can be expressed as
\begin{equation}
    \text{SINR}_n=\frac{\left | \mathbf{w}_n^H\tilde{\mathbf{h}}_n \right |^2 p_n}{\sum^N_{k= 1,k\ne n} \left | \mathbf{w}_k^H\tilde{\mathbf{h}}_k \right |^2 p_k+\left \| \mathbf{w}_n \right \|_2^2\sigma^2  }.\label{sinr}
\end{equation}
\subsection{Computation Task Offloading Model}
The offloading rate of the $n$-th user to FA-BS can be expressed as
\begin{equation}
    R_n = B\log (1+\text{SINR}_n)\label{offloading_rate},
\end{equation}
where $B$ is the bandwidth. When processing a computation task, users have the option to either perform the calculations locally or offload the task to the MEC server for processing. Therefore, these two cases will be described separately below.
\par In the case that computation task is executed locally, the local execution delay of the $n$-th user is
\begin{equation}
     t_n^l = \frac{C_n}{f_n^l},
\end{equation}
where $C_n$ is the size of the computation task of the $n$-th user, and $f_n^l$ is the local computation capacity of the $n$-th user.
\par In the case of computation task is offloaded to the MEC server for calculation, the process of offloading task to the MEC server should be considered, thus the transmission delay of the $n$-th user can be expressed as
\begin{equation}
    t_n^t = \frac{C_n}{R_n}.
\end{equation}
\par The execution delay can be expressed as
\begin{equation}
    t_n^{exe}=\frac{C_n}{\beta_nF^{max} }, 
\end{equation}
where $F^{max}$ represents the maximum execution capacity of the MEC server, $\beta_n\in \left [ 0,1 \right ]$ represents the allocation ratio of the MEC server and $\sum_{n=1}^{N} \beta _n \le 1$. We define $\mathbf{Z} = [\beta _1,\beta _2,\cdots,\beta _N]$ as optimization problem that contains the resource allocation ratio of $N$ MEC servers.
\par If the offloading delay is greater than the local execution delay, the user will choose to execute the task locally. Otherwise, the raw data will be offloaded and processed at the edge server. Since the processing result is far smaller than the offloading data, we focus on the uplink transmission and processing delay of data offloading in this paper. Therefore, the total delay of the $n$-th user is
\begin{equation}
    t_n = \min \left \{ (t_n^t+t_n^{exe}),t_n^l \right \}.
\end{equation}
\subsection{Problem Formulation}
\par The goal of this paper is to minimize the maximum delay among all users by jointly optimizing the APV $\tilde{\mathbf{d}}$, the receive beamforming matrix $\mathbf{W}$, the user transmit power $\mathbf{P}$, and the MEC server computing resource allocation ratio $\mathbf{Z}$. Therefore, the optimization problem can be expressed as
\begin{equation}
    \begin{aligned} 
    \mathcal{P} 1: \min_{\{\tilde{\mathbf{d}}, \mathbf{W}, \mathbf{P},\mathbf{Z}\}} & \max_n \ t_n \\
    \text { s.t. } & \mathrm{C} 1: \mathbf{d}_n\in\mathcal{C}_d,n\in \{ 1,\cdots ,N\}, \\
    & \mathrm{C} 2: tr(\mathbf{w_n} \mathbf{w_n}^H )\le 1,n\in\{1,\cdots ,N\}, \\
    & \mathrm{C} 3: p_n\le p_{max},n\in\{1,\cdots ,N\},\\
    & \mathrm{C} 4: \sum_{n=1}^{N} \beta_iF^{max} \le F^{max},\\
    \end{aligned}
\end{equation}
where Constraint $\mathrm{C} 1$ represents the positioning restriction of the FA ports. Constraint $\mathrm{C} 2$ represents the receive beamforming restriction. Constraint $\mathrm{C} 3$ represents the transmit power of users should not be larger than the maximum transmit power $p_{max}$. Constraint $\mathrm{C} 4$ ensures that the computation capacity of all users on the MEC server does not exceed the maximum computation capacity $F^{max}$. By observing the optimization problem $\mathcal{P} 1$, it is clear that problem $\mathcal{P} 1$ is a non-convex optimization problem with very high dimension. In general, solving these problems effectively poses considerable difficulties. To address this problem, we can estimate accurate channel information through the IBM-CCS and solve part of the high-dimensional optimization through game theory scheme.

\section{IBM-CCS Channel Estimation Scheme}
This section elaborates on the training framework of the proposed IBM-CCS scheme, with particular focus on its suitability for channel estimation in FA scenarios. The detailed network architecture of each component is also presented.
\begin{figure*}[htbp]
    \center{\includegraphics[width=18cm]{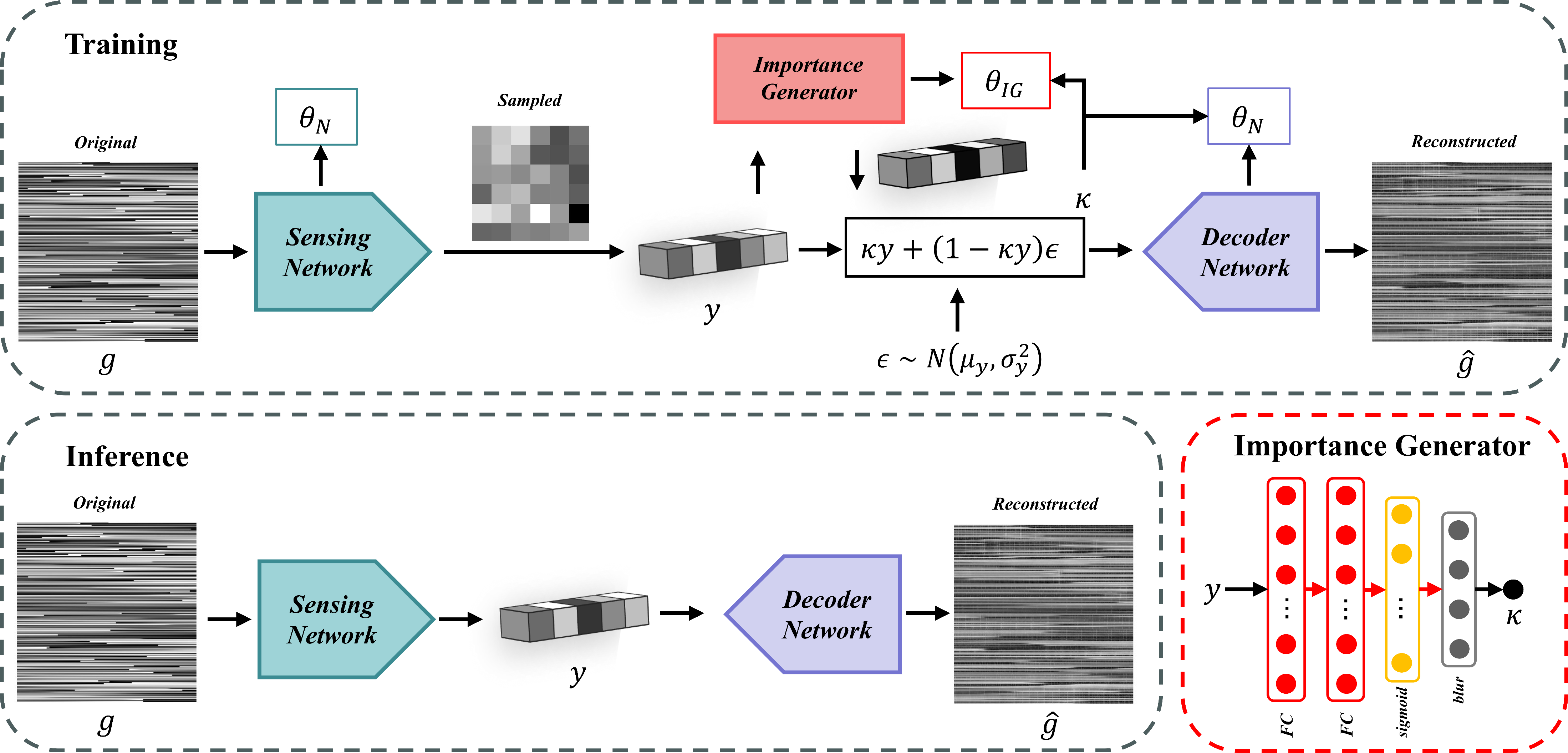}} 
    % 调整宽度
    \caption{The overall scheme of the IBM-CCS.}
    \label{fig:ibm}
\end{figure*}
\subsection{IBM-CCS Training Framework}
The overall scheme of IBM-CCS channel estimation is shown in Fig.~\ref{fig:ibm}. In typical CS-based reconstruction pipelines, a sensing network $S(\cdot)$ captures low-dimensional measurements $y$ from a high-dimensional input, while a decoder network $D(\cdot)$ attempts to reconstruct the original input from $y$. In the context of FAS, the complex-valued spatial channel responses that are shaped by the unique reconfigurability and location-dependent fading characteristics of FAs are first transformed into two-dimensional grayscale images to facilitate learning-based processing. These channel images reflect the spatial diversity and dynamic fading environments inherent to FA deployments. 

Due to the continuous movement and adaptive positioning of FA, the resulting channel distributions exhibit high sparsity and localized structures, making them particularly well-suited for CS-based methods. However, this also means that preserving the essential information under aggressive compression is more critical than in conventional static antenna systems. To address this, the IBM-CCS scheme integrates the Information Bottleneck principle into the CS pipeline, guiding the sensing network to extract the most relevant latent representations for accurate reconstruction, while discarding redundant variations that do not significantly impact estimation fidelity. This targeted information preservation is especially beneficial in FAS, where the spatial coherence and channel statistics can change rapidly with slight antenna displacements. As a result, IBM-CCS provides a robust and efficient estimation framework that adapts naturally to the fluid and non-stationary properties of FA channels.

% In this paper, the complex channel coefficients are represented as a two-dimensional grayscale image as the input of the original image. The input image is the result of a specific sampling of the original image, and the sensing network generates the measurement value by processing the sampled image. Since the dimension of the measurement value is much lower than the dimension of the original image, it is inevitable that some degree of information will be lost. For this reason, it is particularly important to retain the main information of the original image as much as possible in the low-dimensional measurement value during the compression process.
To address the aforementioned challenge of effectively preserving critical information in the low-dimensional measurements, we propose an importance generator $IG(\cdot)$ that consists of two consecutive fully connected layers, sigmoid function, and blur. Its input is the measurement value $y$ of the compressed input image $g$, and its output is the importance value $\kappa \in \mathbb{R} ^M$. To estimate importance value of measurements, we hope to reduce the information in $y$ by adding noise. As information bottleneck, a linear interpolation between $y$ and noise is applied as
\begin{equation}
    Z = \kappa y+(\mathbf{1}-\kappa y)\epsilon ,\label{intermediate}
\end{equation}
where $\kappa=IG(y)$, $\epsilon\sim\mathcal{N}(\mu_y,\sigma_y^2)$ is the noise with a mean of $\mu_y$ and a variance of $\sigma ^2_y$, and $\mathbf{1}$ are the vector form of ones. $Z$ is the intermediate variable that will be forwarded to the decoder $D(\cdot)$ to output the reconstructed image $\hat{g}$.
To train the importance generator that aims at maximizing the information shared by the intermediate variable $Z$ and the reconstruct target $g$ while reducing the information shared by $Z$ and input variable $y$, which is formulated as
\begin{equation}
    \max I[g;Z]- \eta I[y;Z],\label{ib}
\end{equation}
where $I[g;Z]$ represents the mutual information and $\eta$ is used to balance the trade-off between estimating target $g$ and feature information of input $y$. $I[y;Z]$ represents the amount of information of $y$ in the intermediate variable $Z$. According to \cite{ib}, the upperbound of $I[y;Z]$ can be formulated as
\begin{equation}
      L_I = \mathbb{E} _y\left [ D_{KL}\left [ P(Z|y) || \mathcal{N}(\mu_y,\sigma _y) \right ]  \right ], 
\end{equation}
where $P(Z|y)$ denote the respective probability distributions. $D_{KL}\left [ \cdot||\cdot \right ]$ denote the Kullback-Leible (KL) divergence between two normal distributions. We can know that $L_I \ge I[y;Z]$, which is considered as a part of reconstruction loss of importance generator. The main part of reconstruction loss comes from the first term of (\ref{ib}), since the recovery performance depends on the correlation between intermediate variable $Z$ and reconstruction target $g$. Therefore, the reconstruction loss of importance generator can be formulated as
\begin{equation}
    L_{IG} = L_R + \eta L_I,\label{eq:loss_ig}
\end{equation}
where $L_R = \left \| g-\hat{g}  \right \|_2^2 $. The trained importance generator controls the weight of the decoder network input information through the output variable $\kappa$. The weight of important elements of $y$ becomes closer to 1, and the weight of relatively unimportant elements becomes smaller, which greatly improves the reconstruction performance.
\par To train sensing network and decoder network, we design loss function of sensing and decoder networks based on importance generator for further training, which can be formulated as
\begin{equation}
    L_N = L_R+ \gamma\left \| \mathbf{1}-\kappa   \right \|_2^2,\label{eq:loss_sd}
\end{equation}
where $\gamma$ is the regularization parameter. The variable $\kappa$ can adjust the sensing and decoder networks, significantly improving the reconstruction performance.
\begin{algorithm}
    \caption{IBM-CCS Training Algorithm}\label{alg:ibm}
    \begin{algorithmic}[1]
    \STATE $\mathbf{Initialize:}$ Network parameters $\theta_{IG}$ of importance generator, network parameters $\theta_{N}$ of sensing network and decoder network.
    \WHILE{not converged}
        \FOR{Time slot $t = 1,2,\ldots,N_t$} 
            \STATE Initialize the channel information.
            \STATE Transform channel information into grayscale image $g$
            \STATE Extract patch $g$.
            \STATE Compress information $y = S(g)$.
            \STATE Train importance generator $IG(\cdot)$ with (\ref{eq:loss_ig}), and update the network parameters $\theta_{IG}$.
            \STATE Get importance $\kappa$ with $\kappa = IG(y)$.
            \STATE Train sensing network $S(\cdot)$ and decoder network $D(\cdot)$ with (\ref{eq:loss_sd}), and update the network parameters $\theta_{N}$.
        \ENDFOR
    \ENDWHILE
    \end{algorithmic}
\end{algorithm}
\subsection{Sensing Network}
\begin{figure}[htbp]
    \centering
    \includegraphics[width=0.48\textwidth] {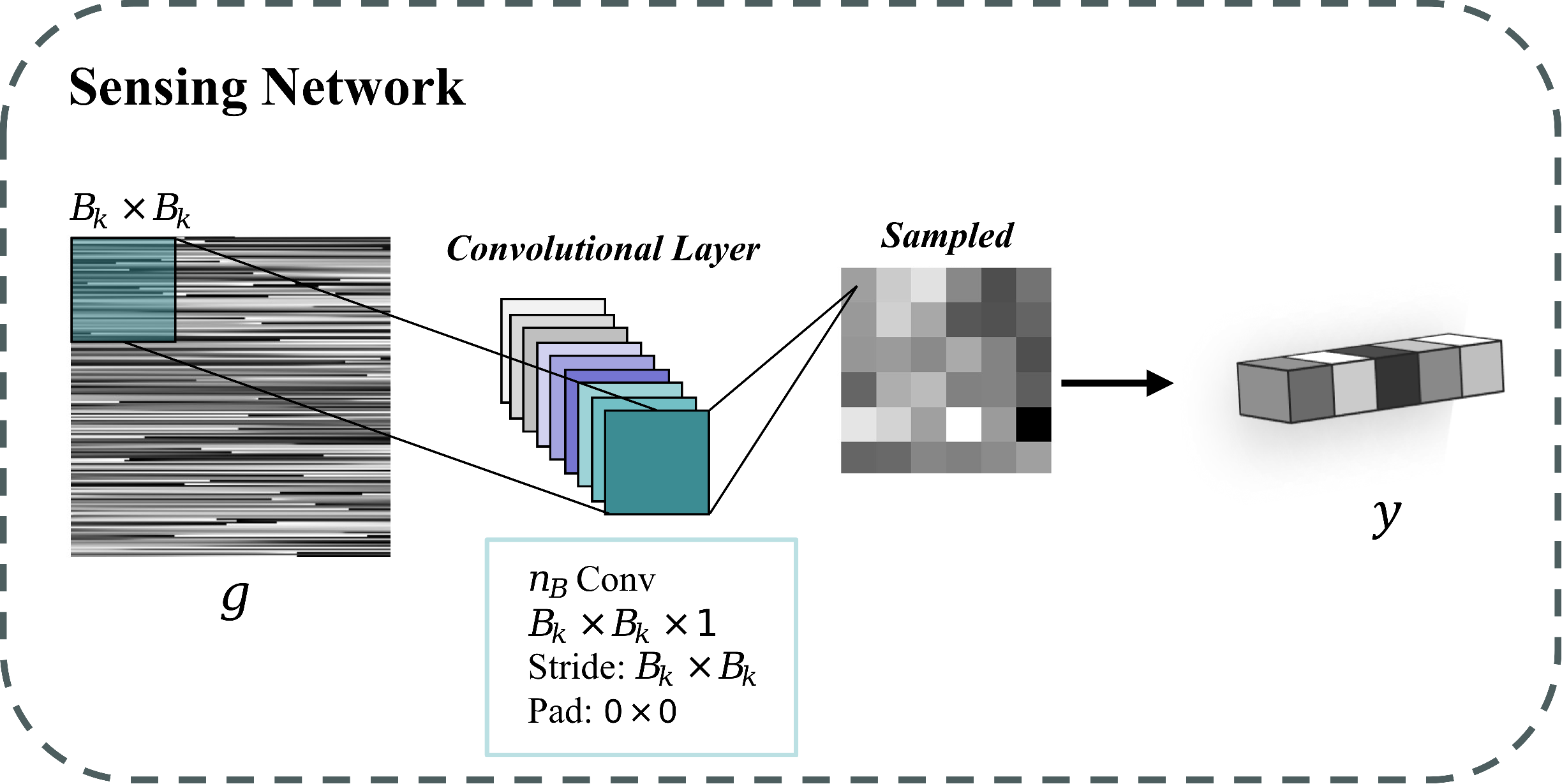} % 调整宽度
    \caption{The structure of the sensing network.}
    \label{fig:SN}
\end{figure}
Fig.~\ref{fig:SN} shows the basic structure of the sensing network, which uses block-based compressed sampling (BCS) to learn the sampling matrix and obtain channel compressed sensing(CCS) measurements. In BCS, the channel matrix is partitioned into non-overlapping $B_k \times B_k \times 1$ blocks. The CCS measurements are obtained via a sampling matrix $\Phi_B$ of size $n_B \times B_k^2$, mathematically expressed as $ y = \Phi_B g$. This operation is modeled as a convolutional layer where each matrix row corresponds to a $B_k \times B_k\times 1$ kernel, producing one measurement per kernel. For a sampling ratio $r$, the layer contains $n_B = rB_k^2$ kernels with $B_k \times B_k$ strides and zero bias, ensuring non-overlapping coverage. The sampling network $S(g)$ is formally defined as
\begin{equation}
    y=S(g)=W_s \ast g,
    \label{eq:sg}
\end{equation}
% Consistent with prior works \cite{mun2011residual}, we implement $B_k=32$. At $r=0.1$, this yields $n_B = 0.1 \times 32^2 = 102$ filters. 
where $\ast$ denotes convolution, $W_s$ represents $n_B$ learnable kernels of size $B_k \times B_k$, and no activation or bias is applied. Each output feature map column corresponds to the measurement of a specific block. During training, $W_s$ can be adaptively learned to preserve the structural information in the measurements, which is particularly valuable for FAS where the spatial coherence and fading statistics of the channel may vary rapidly with antenna displacement. Once trained, $W_s$ can be employed as an efficient encoder to perform real-time CS-based acquisition of FA channel measurements.
\subsection{Decoder Network}
\begin{figure*}[htbp]
    \center{\includegraphics[width=18cm]{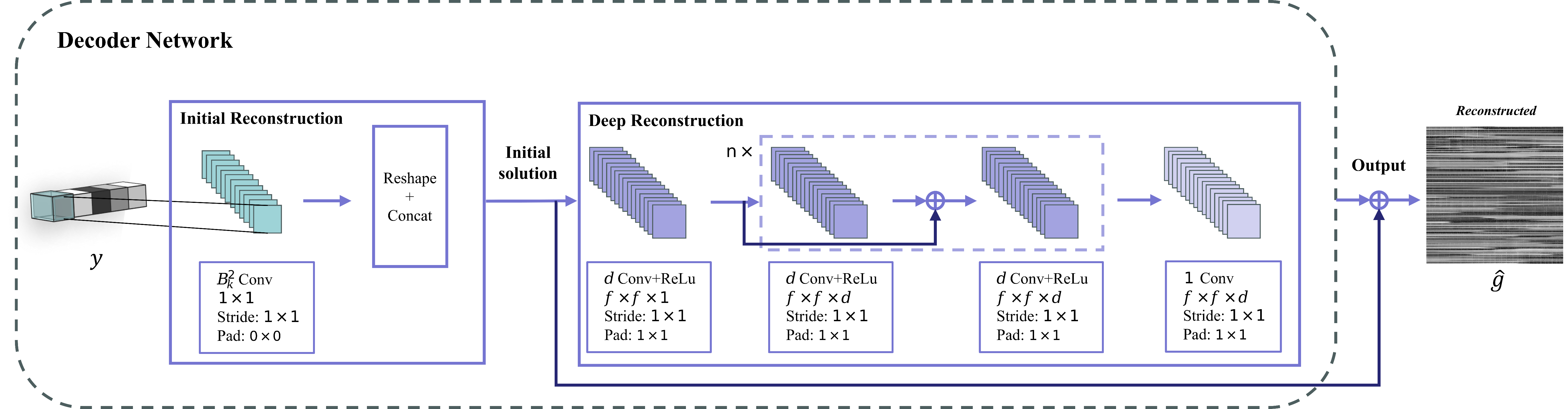}}
     % 调整宽度
    \caption{The structure of the decoder network.}
    \label{fig:DN}
\end{figure*}
To achieve accurate channel recovery from compressed measurements, we design a hierarchical reconstruction architecture comprising two complementary modules, the initial reconstruction module and the deep reconstruction module, as illustrated in Fig.~\ref{fig:DN}. The initial module performs coarse recovery, providing a structural approximation of the channel, while the deep module refines this estimation by extracting and enhancing high-level nonlinear features. Given a CCS measurement vector $y$, the reconstruction process is defined as
\begin{equation}
    D(y)=D_p(I(y)),
    \label{eq:recon_process}
\end{equation}
where $D(\cdot)$ represents the decoder reconstruction network, $I(\cdot)$ and $D_p(\cdot)$ represent the initial reconstruction network and the deep reconstruction network respectively.
\subsubsection{Initial Reconstruction Network}
In traditional BCS, the pseudo-inverse of the sampling matrix $\tilde{\Phi_B}$ is usually employed to generate the initial reconstruction from the CS measurements \cite{zhang2014group}. Instead of relying on this fixed transformation, we introduce a learnable convolutional layer to adaptively reconstruct spatial channel blocks.  The initial reconstruction layer $\tilde{I}(\cdot)$ can be mathematically expressed as
\begin{equation}
    \tilde{I}(y) = W_{int}\ast y,
    \label{eq:init_recon_process}
\end{equation}
where $W_{int}$ is a set of trainable kernels. Specifically, since the output of the sampling network is a $1\times1\times n_B$ vector, the initial reconstruction layer $\tilde{I}(\cdot)$ utilizes $B_k^2$ kernels (size $1\times1\times n_B$) with $1\times1$ strides and no bias. Each kernel reconstructs a $B_k \times B_k \times 1$ channel block from its corresponding measurement vector.

% However, the reconstructed output of each block is a $1 \times 1 \times B_k^2$ vector. To improve the traditional method of utilizing reshaped and concatenated block vectors to form the initial channel, we implement a composite layer integrating both reshaping ($\chi$) and concatenation ($\pi$) functions. Each reconstructed $1 \times 1 \times B_k^2$ vector is transformed via $\chi(\cdot)$ into a $B_k\times B_k \times 1$ block, which is then spatially reassembled by $\pi(\cdot)$ into a full-channel reconstruction. This process can be expressed as
To reconstruct the entire channel matrix, each block output which is a size of $1 \times 1 \times B_k^2$ vector is reshaped and then spatially arranged into its original location using a composite operation $I(y)$ defined by
\begin{equation}
    \tilde{g}=I(y)=\pi \left(\begin{array}{ccc}
    \chi\left(\tilde{I}_{11}(y)\right) & \cdots & \chi\left(\tilde{I}_{1 w}(y)\right) \\
\vdots & \ddots & \vdots \\
\chi \left(\tilde{I}_{h 1}(y)\right) & \cdots & \chi\left(\tilde{I}_{h w}(y)\right)
\end{array}\right)  , 
\label{eq:reshape}
\end{equation}
where $\chi(\cdot)$ reshapes a $1 \times 1 \times B_k^2$ vector into a $B_k \times B_k \times 1$ block, and $\pi(\cdot)$ reassembles all blocks into the full reconstructed image. The holistic processing approach allows the model to consider both intra-block structure and inter-block spatial correlations, which is especially critical in FAS where the spatial channel distribution is non-stationary and dynamically coupled to antenna position.
% where $\tilde{I}_{ij}(y)$ denotes a $1\times 1\times B_k^2$ vector of block at spatial coordinates $(i,j)$, with $h\times w$ representing the block grid dimensions. This architecture enables holistic optimization across inter-block correlations rather than isolated block processing, effectively leveraging both intra-block details and inter-block contextual relationships. 

\subsubsection{Deep Reconstruction Network}

To further enhance the reconstruction quality, we employ a deep residual reconstruction architecture that extracts nonlinear spatial features and compensates for lost details. The framework comprises three sequential components, namely feature extraction, non-linear transformation, and feature aggregation.

% The initial stage utilizes convolutional operations to extract high-dimensional feature representations from local receptive fields. This module consists of an initial convolutional layer followed by a nonlinear activation layer. The convolution operation processes the initial reconstruction output $\tilde{g}$ derived from Eq. (\ref{eq:reshape}) using $d$ trainable kernels with dimensions $f\times f\times 1$. 
In the feature extraction stage, convolutional layers process the initial reconstruction $\tilde{g}$ using $d$ kernels of size $f \times f \times 1$, followed by a ReLU activation. Mathematically, this operation can be expressed as 
\begin{equation}
    D_e(\tilde{g}) = \text{ReLU}(W_e \ast \tilde{g} + B_e),
\end{equation}
where $W_e$ and $B_e$ denote the weights and biases, respectively. The rectified linear unit $\text{ReLU}(x)=max(0,x)$ is employed as the primary activation function. This produces a high-dimensional feature map capturing localized patterns.

Next, the deep reconstruction architecture employs an alternating cascade of residual blocks with convolutional and activation layers following high-dimensional feature extraction, effectively expanding network nonlinearity and model expressiveness. This non-linear transformation operation is formulated as
\begin{equation}
    D_{m1}^i(\tilde{g}) = \text{ReLU}(D_{m2}^{i-1}(\tilde{g})+W_{m1}^i \ast D_{m2}^{i-1}(\tilde{g}) + B_{m1}^i ),
    \label{eq:residual}
\end{equation}
\begin{equation}
    D_{m2}^i(\tilde{g}) = \text{ReLU}(W_{m2}^i \ast D_{m1}^{i}(\tilde{g}) + B_{m2}^i ),
\end{equation}
where $i \in \{ 1,2,\cdots,n\}$, (\ref{eq:residual}) defines the residual block structure incorporating short skip connections between input and output features. $W_{m1}^i$ and $W_{m2}^i$ represent the $d$ trainable kernels of size $f\times f \times d$, $B_{m1}^i$ and $B_{m2}^i$ denote the bias of size $d\times 1$.

After $n$ residual blocks, a final feature aggregation layer reconstructs the refined output, mathematically formulated as
\begin{equation}
    D_{a}(\tilde{g})=W_a \ast D_{m2}^n(\tilde{g})+B_a,
\end{equation}
where $W_{a}$ represent the $1$ trainable kernel of size $f\times f \times d$, and $B_{a}$ denote the bias of size $1 \times 1$. To enhance convergence efficiency, we incorporate long-range skip connections \cite{kim2016accurate} between initial reconstruction output $\tilde{g}$ and deep reconstruction network output $D_{a}(\tilde{g})$. The final reconstructed image is therefore obtained through
\begin{equation}
    D_{p}(\tilde{g})=\tilde{g}+D_{a}(\tilde{g}).
\end{equation}

The IBM-CCS framework enables highly accurate and low-dimensional channel estimation in FAS, which are characterized by spatially adaptive and dynamic signal propagation environments. However, accurate channel knowledge alone is not sufficient for optimizing overall system performance. In FA-assisted MEC network, how to intelligently offload computation tasks under constraints such as limited power and fluctuating channels becomes critical. 
\section{Game Theory Scheme for Power Design}
User power allocation fundamentally governs computation offloading efficiency through dual effects observed in (\ref{sinr}), which increasing transmit power improves channel gain yet induces inter-user interference, while DRL complexity grows with optimization variables. Notably, the high-dimensional nature of $\mathcal{P}1$ amplifies this complexity. By employing game theory to allocate user transmit power effectively, we can enhance computation offloading while simultaneously reducing the dimensionality of the optimization variables in $\mathcal{P}1$.
\par Non-cooperative game theory is an efficient method to design power allocation. In this section, we use equal power design as the initial power distribution and then perform non-cooperative games on the transmit power of users. During the game, a pricing mechanism is adopted to adjust the transmit power of each user, thus achieving interference coordination. FA-BS can be regarded as a participant, and the process of allocating transmit power from users to FA-BS can be viewed as a game, which can be denoted by
$G=\left[\mathbf{U},{\mathbf{P}},\left\{f_{ n}^{c}(\cdot)\right\}\right]$. Specifically,
$\mathbf{U} = \{u_1,\cdots ,u_{N}\}$ represents the players, i.e., users participating in the communication. $\mathbf{P}=\{p_n\in[0,p_n^{(0)}]\mid \forall n\in \{1,\cdots,N \} \}$ denotes the set of available strategies for players, where $p_n^{(0)}$ denotes the initial power allocated to the $n$-th user.  $f_{ n}^{c}(\cdot)$ represents the net utility function of the $n$-th user. More
specifically, $f_{ n}^{c}(\cdot)$ can be expressed as
\begin{equation}
    f_{ n}^{c}(p_n,{p_n}' )=f_{ n}(p_n,{p_n}' )-c_n(p_n,{p_n}' ),
\end{equation}
where ${p_n}'$ is the interference power from other users except the $n$-th user to FA-BS. In addition,  $f_{ n}(p_n,{p_n}' )$ and $c_n(p_n,{p_n}')$ represent the utility function and pricing function, respectively. In this paper, they are defined as
\begin{equation}
    f_{ n}(p_n,{p_n}' )=R_n\label{utility},
\end{equation}
\begin{equation}
    c_n(p_n,{p_n}' )=\xi_n p_n \label{pricing},
\end{equation}
where $\xi_n$ is the pricing factor.
\par Computing the first derivative of (\ref{utility}), we have
\begin{equation}
    \frac{\partial f_{ n}^{c}(p_n,{p_n}')}{\partial p_n} = \frac{\phi_n }{\ln2(\tilde{I}_n+\delta_n ^2 +\phi_n p_n )  } - \xi_n ,\label{eq:order}
\end{equation}
where $\tilde{I}_n = \sum^{N}_{k\ne n} \left | \mathbf{w}_k^H\mathbf{h}_k \right |^2 p_k$, $\phi_n=\left | \mathbf{w}_n^H\mathbf{h}_n \right |^2$, and $\delta_n^2 = \left \| \mathbf{w}_n \right \|_2^2\sigma^2$. Let (\ref{eq:order}) equals to 0 to obtain the optimal power response as

\begin{equation}
    p_n=\frac{1}{\xi_n\ln2} -\frac{\tilde{I}_n + \delta _n^2 }{\phi_n}. \label{power}
\end{equation}

\par Since $p_n \in \left [ 0,p_n^{(0)} \right ]$, the upper bound $\xi_n^{max}$ and the lower bound $\xi_n^{min}$ of $\xi_n$ can be calculated as
\begin{equation}
    \xi_n^{max}=\frac{\phi_n}{\ln2(\tilde{I}_n+\delta _n^2)},\label{max}
\end{equation}

\begin{equation}
    \xi_n^{min}=\frac{\phi_n}{\ln2(p_n^{(0)} \phi_n + \tilde{I}_n + \delta_n ^2)}.
\end{equation}
\par It can be obtained that $\frac{\xi_n-\xi_n^{min}}{\xi_n^{max}-\xi_n^{min}}\in\left [ 0,1 \right ]$. To design a new price factor $\lambda$ to make sure $p_n \in \left [ 0,p_n^{(0)} \right ]$, let 
\begin{equation}
    \Gamma = e^{-\lambda(\phi_n\times\vartheta)^\varphi}=\frac{\xi_n-\xi_n^{min}}{\xi_n^{max}-\xi_n^{min}},\label{gamma}
\end{equation}
where $\varphi$ is a positive constant. $\vartheta$ is used to adjust the order of magnitude, which is defined as $\vartheta=\frac{\nu }{\max\{\phi_n\}} $, where $\nu$ is a positive constant and $\max\{\phi_n\}$ represents the maximum channel gain between all users and FA-BS. In addition, $\lambda$ is used as a new pricing factor, and $\lambda > 0$. Through calculation, it can be observed that
\begin{equation}
    \xi_n = \Gamma  \xi_n^{max}+(1-\Gamma)\xi_n^{min}.\label{factor}
\end{equation}
\par Observing from (\ref{gamma}) and (\ref{factor}) that $\xi_n$ is adjusted according to $\phi_n$ for different users, while $\lambda$ is the same for all users. From (\ref{gamma}), we can observe that users with larger $\phi_n$ will make $\xi_n$ closer to $\xi_n^{min}$. Therefore, users with larger $\phi_n$ will match smaller $\xi_n$ to reduce the price $c_n(p_n,{p_n}' )$ in (\ref{pricing}) and achieve a larger net utility function value in (\ref{utility}).
\par Substituting (\ref{max})-(\ref{factor}) into (\ref{power}), the power iteration formula can be obtained by
\begin{equation}
    p_n^{\tau+1 }=\frac{\left [ 1- e^{-\lambda(\phi_n\vartheta)^\varphi}\right ](\tilde{I}_n^{(\tau)}+\delta_n ^2) p_n^{(0)} }{\tilde{I}_n^{(\tau)}+\delta_n ^2+e^{-\lambda(\phi_n\vartheta)^\varphi} p_n^{(0)} \phi_n },\label{power_iteration}
\end{equation}
where $\tau$ represents the number of iterations.
\par Based on game theory, the original optimization problem $\mathcal{P} 1$ can be reconstructed to $\mathcal{P} 2$ as 
\begin{equation}
    \begin{aligned}
        \mathcal{P} 2: \min_{\{\tilde{\mathbf{d}}, \mathbf{W}, \lambda,\mathbf{Z}\}} & \max_n t_n \\
        \text { s.t. } 
        & \mathrm{C} {3}' : \lambda >0,\\
        &  \mathrm{C} 1,\ \mathrm{C} 2,\ \mathrm{C} 4,\\
    \end{aligned}
\end{equation}
where $\mathrm{C} {3}'$ constrains the pricing factor $\lambda$ to be a positive number. In this way, the dimension of transmit power design is reduced from $N$ dimensions to 1, which reduces the complexity of the problem. However, the optimization problem is still quite complex and non-convex. Therefore, the DRL method is introduced below to solve this problem.

\section{HiTDMA Based Offloading Scheme}

Building upon the IBM-CCS channel estimation framework and the game-theoretic power allocation strategy designed for dimension reduction, this section further explores the decision-making layer of the FA-assisted MEC offloading system. Given that accurate channel state information and rational power distribution lay the foundation for efficient offloading, it is now crucial to address how users and the BS collaboratively make dynamic and efficient offloading decisions in real time.

To this end, the network training process is formulated as a Markov Decision Process (MDP) problem, which captures the sequential and uncertain nature of FA-assisted MEC offloading system. DRL proves to be particularly effective in solving such MDPs through iterative interactions between the agent and the environment. To address the optimization problems $\mathcal{P} 2$, we adopt a MADRL framework. Specifically, the agents are categorized into two types, DUAs and TBAs. The DUAs are responsible for solving the port selection problem for the FA associated with each user. Meanwhile, the TBAs handle joint optimization tasks, including beamforming, the allocation of MEC server computation resources, and the adjustment of pricing factor. To avoid convergence to suboptimal solutions, agents within each category share information, enabling cooperative learning and improved decision-making.

\begin{figure}[htbp]
    \centering
    \includegraphics[width=0.48\textwidth] {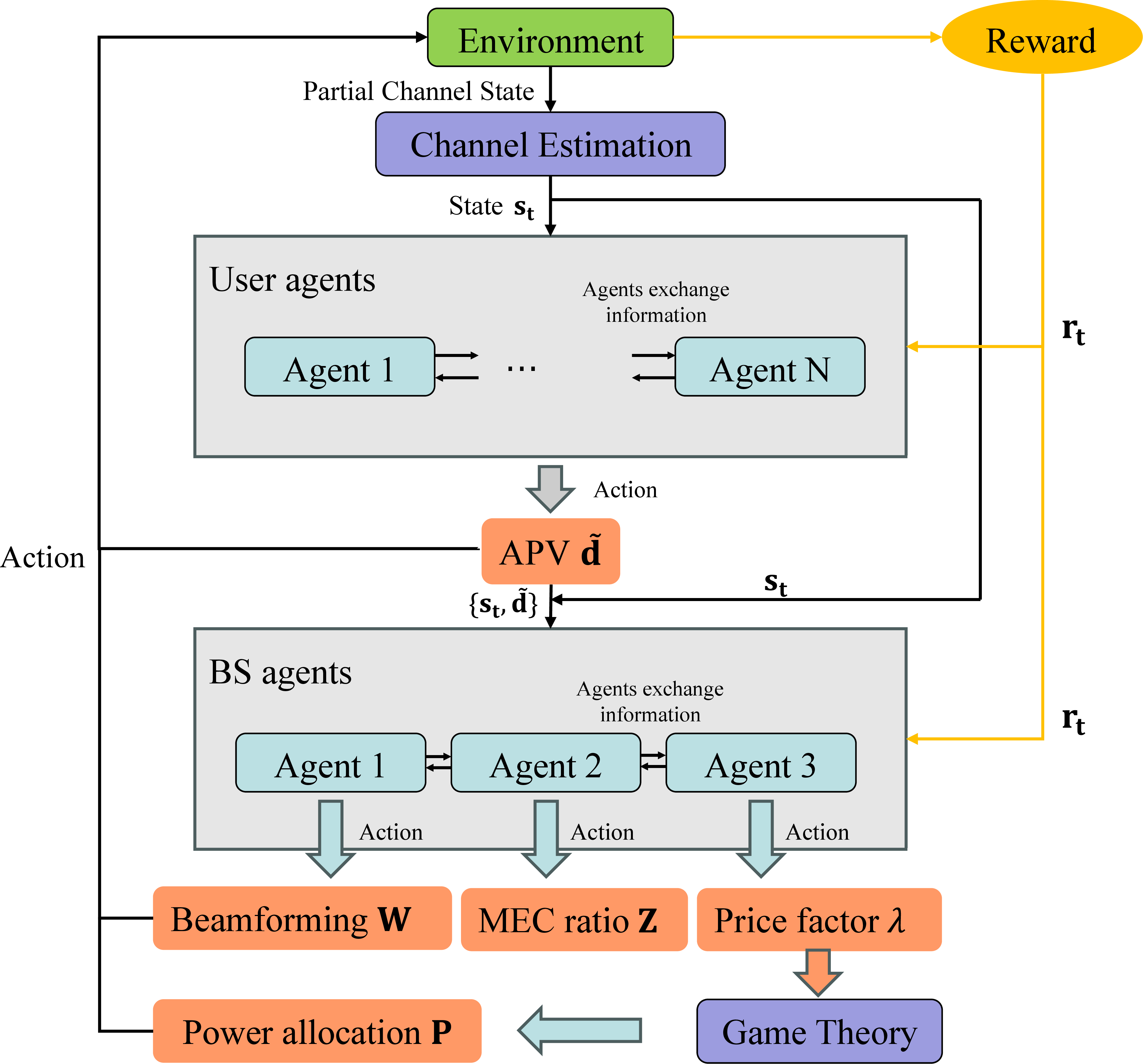} % 调整宽度
    \caption{The architecture of the HiTDMA-based computation offloading.}
    \label{fig:drl}
\end{figure}

In the formulation of this MDP problem, the tuple $(S,A,P,R)$ is defined as follows. $S$ represents the set of all possible states of the environment, encompassing factors such as user locations, channel information, offloading rates, and the port locations of FA. $A$ denotes the action space available to the agents, which includes decisions on FA port locations, beamforming matrix, allocation ratio of MEC server computing resource, and pricing factors. $P = \left \{ p( s_{t+1}\mid s_t,a_t) \right \} $ represents the state transition probabilities, while $R$ is the set of rewards received by the agents, which are functions of the states and actions. Moreover, the rewards are closely tied to the optimization objectives and constraints of the system.
The dynamic process of the MDP is described as follows. At time $t$, each agent observes the current state $s_t$ from the environment and selects an action $a_t$. Subsequently, the environment transitions to a new state $s_{t+1}$ based on the action taken, and the agent receives a corresponding reward $r_t$.

As shown in Fig. \ref{fig:drl}, we design a HiTDMA based computation offloading structure. Given that the FA port selection problem is a discrete optimization problem, we adopt the DUA based on Dueling Double Deep Q Network~(D3QN), which is well-suited for solving discrete problems. Conversely, for the continuous optimization tasks tackled by the TBA, which is based on the twin Delayed Deep Deterministic Policy Gradient~(TD3). In the subsequent sections, we first introduce the DUA framework. This is followed by an explanation of the TBA framework. Finally, we present the HiTDMA that integrates these frameworks, with details elaborated in the following discussion.
\subsection{Dueling-based User Agent}
\begin{figure}[htbp]
    \centering
    \includegraphics[width=0.47\textwidth] {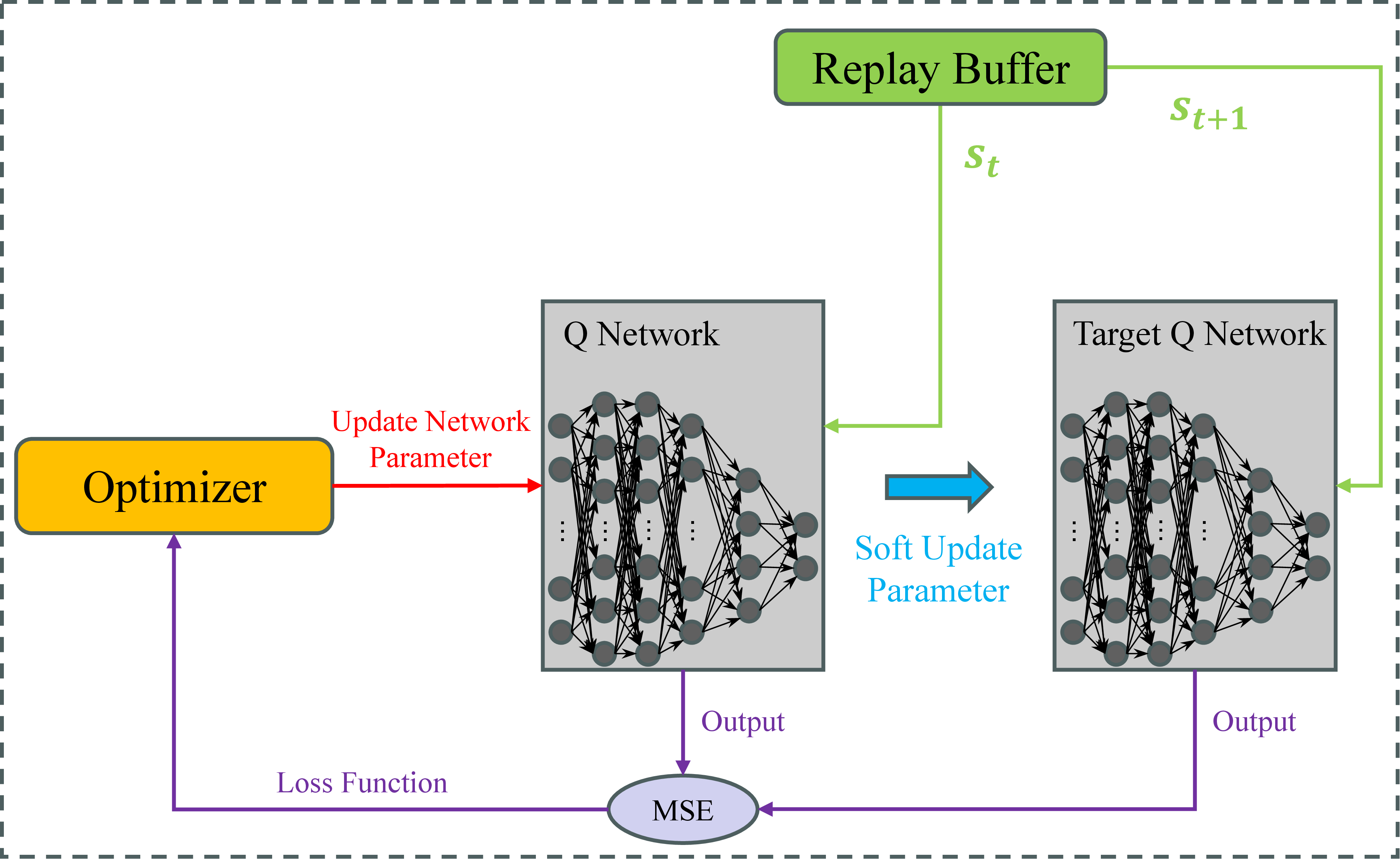} % 调整宽度
    \caption{The structure of DUA.}
    \label{fig:d3qn}
\end{figure}
    \subsubsection{State Space}In the $t$-th training time slot, the state of the DUAs only includes the current small-scale channel information and the location of all users, and the total delay of the users at the previous time. Therefore, the state of the DUAs can be expressed as
    \begin{equation}
        s_t^u=\left [ \mathbf{g}^t,loc^t,\mathbf{t}^{t-1}  \right ] ,
    \end{equation}
    where $\mathbf{g}^t=\left [ \mathbf{g}_1^{a,t},\mathbf{g}_2^{a,t},\cdots,\mathbf{g}_N^{a,t} \right ] $ and $\mathbf{g}_n^a $ represents the whole channel information of all preset ports to the $n$-th user at FA. $loc^t$ is the position information of the users at the time slot $t$.
    
\subsubsection{Action Space}We define the APV on the FA as action according to the state information at the current time slot, which can be expressed as
\begin{equation}
    a_{t,n}^u= \mathbf{d}_n. 
\end{equation}

    \subsubsection{Reward Design}The design of the reward is crucial to the convergence speed and performance of the model. According to the optimization problem $\mathcal{P} 1$, our goal is to minimize the maximum user total delay. Therefore, the reward design of the DUAs is closely related to the delay. Based on this, we design a reward related to time delay, which can be expressed as
\begin{equation}
    r_t=\left\{\begin{array}{ll}
\delta_{1}, & \text { if } T_{s} \leq t_{1}, \\
\frac{\delta_{1}\left(t_{2}-T_{s}\right)}{t_{2}-t_{1}}, & \text { if } t_{1}<T_{s} \leq t_{2}, \\
0, & \text { otherwise},
\end{array}\right.\label{eq:reward design}
\end{equation}
where $t_{1}$ and $t_{2}$ represent preset delay thresholds with $t_{1}<t_{2}$, $t_{1}$ is the desired delay target for service quality guarantee, while $t_{2}$ defines the maximum tolerable delay beyond which no reward is granted. $T_{s} = {\max_n} t_n$ is the maximum delay for all users.
    
    \subsubsection{D3QN Algorithm}To determine the optimal APV, the DUA implements the D3QN that synergistically integrates the architectural merits of Double Deep Q-Network (DDQN) and Dueling Deep Q-Network. Its structure is shown in Fig. \ref{fig:d3qn}. This integration addresses two critical challenges in reinforcement learning, environmental state dependency and value estimation accuracy. 

    The conventional Deep Q-Network (DQN) framework exhibits inherent limitations in value estimation due to its action-centric Q-function formulation. While the baseline architecture computes Q-values solely based on selected actions, empirical studies reveal significant environmental state dependencies in specific tasks. This action-induced value coupling limitation frequently results in inefficient policy learning and low speed of convergence during model training.

    To address this critical challenge in value function approximation, the dueling DQN introduces an innovative structural decomposition mechanism. Through decoupled representation learning, the Q-function is analytically separated into state-value function $V(s_t^u;\theta, \psi)$ and action-advantage function $A(s_t^u,a_t^u;\theta, \phi)$ components, formally expressed as
\begin{equation}\label{rt}
Q(s_t^u,a_t^u;\theta, \psi, \phi)=V(s_t^u;\theta, \psi)+A(s_t^u,a_t^u;\theta, \phi),
\end{equation}
where $\theta$ is the parameter that both the value function and the advantage function have. $\psi$ is the parameter that only the value function has and $\phi$ is the parameter that only the advantage function has.

In addition, DDQN solves the problem of the overestimation of Q value caused by traditional DQN using a single neural network for both action selection and target valuation through a dual network architecture. The action selection mechanism adopts
\begin{equation}\label{rt2}
a_t^*=\arg\max \limits_{a^u} Q(s_{t+1}^u,a_t^u;\theta).
\end{equation}

The objective function of the DDQN is
\begin{equation}\label{rt3}
y_t^{\text{DDQN}}=r_{t+1}+ \gamma^* Q(s_{t+1}^u,a_t^*;\theta^-),
\end{equation}
where $\theta$ and $\theta^-$ are the training and target network parameters. The above equation can be rewritten as $y_t^{\text{DDQN}}=r_{t+1}+ \gamma^* Q(s_{t+1}^u,\arg\max \limits_{a^u} Q(s_{t+1}^u,a_t^u;\theta);\theta^-)$.

Therefore, the D3QN algorithm achieves enhanced learning efficacy through the amalgamation of the method of reducing overestimation in DDQN and the method of dividing the Q-function in Dueling DQN. The objective expression of D3QN is similar to the DDQN, which can be expressed as
\begin{equation}\label{rt5}
y_t^{\text{D3QN}}=r_{t+1}+ \gamma^* Q(s_{t+1}^u,\arg\max \limits_{a^u} Q(s_{t+1}^u,a_t^u;\theta);\theta^-).
\end{equation}

The loss function of the D3QN can be expressed as
\begin{equation}\label{rt66}
loss_{\text{D3QN}} (\theta)=[y_t^{\text{D3QN}}-Q(s_t^u,a_t^u;\theta)]^2.
\end{equation}

The expression of the D3QN when updating the target network parameter $\theta^-$ is
\begin{equation} \label{eq10}
\theta^- = \tau_2 \theta + (1- \tau_2) \theta^-,
\end{equation}
where $\tau_2 \in (0,1)$ is the update weight parameter of the D3QN target network, which indicates that the target network retains the previous target network parameters with a weight of $1-\tau_2$ and updates the parameters of the training network to the target network with a weight of $\tau_2$.

\subsection{Twin-critic-based Base station Agent}
\begin{figure}[htbp]
    \centering
    \includegraphics[width=0.5\textwidth] {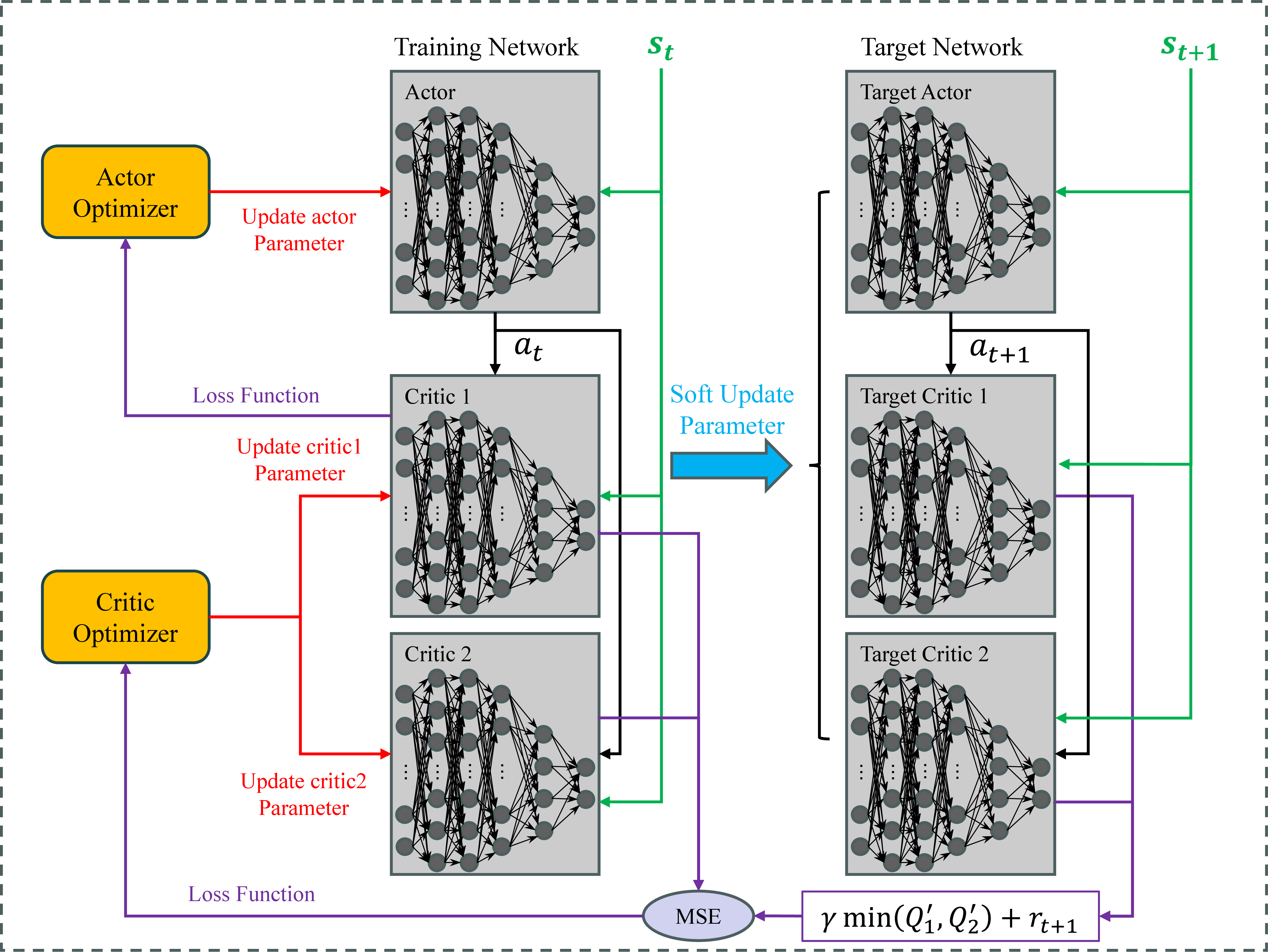} % 调整宽度
    \caption{The structure of TBA.}
    \label{fig:td3}
\end{figure}
    \subsubsection{State Space}The state of the TBAs includes not only the channel information and location of all current users, as well as the total delay of the users at the previous moment but also the FA ports location information which is output by the DUAs. Therefore, the state of the TBAs can be expressed as
\begin{equation}
    s_t^b=\left [ s_t^u,\tilde{\mathbf{d}}   \right ]. \label{eq:base_state}
\end{equation}

    \subsubsection{Action Space}The action output of the TBAs includes the receive beamforming matrix $\mathbf{W}$, the price factor $\lambda$, and the MEC server resource allocation vector $\mathbf{Z}$, where the beamforming contains real and imaginary parts, i.e., $\text{Re}\{\mathbf{W}\}$ and $\text{Im}\{\mathbf{W}\}$. Therefore, the action of the TBAs can be expressed as
\begin{equation}
    a_t^b=\left [ \text{Re}\{\mathbf{W}\}, \text{Im}\{\mathbf{W}\},\lambda,\mathbf{Z} \right ]. 
\end{equation}
    \subsubsection{Reward Design}In this paper, the common goal of the TBAs and the DUAs is to minimize the maximum total delay of the users. Therefore, the reward of the TBAs is the same as the reward of the DUAs.
    \subsubsection{TD3 Algorithm}
    The TBA decision-making framework employs TD3 for joint optimization of beamforming matrix, pricing factor, and MEC server resource allocation ratio. As an enhanced variant of Deep Deterministic Policy Gradient (DDPG), TD3 maintains the core architecture of deterministic policy $\mu(s; \theta_{\mu})$ and Q-value approximators $Q(s, a, \theta_{Q_1})$. Its structure is shown in Fig. \ref{fig:td3}. The DDPG foundation contains four core components, Actor/critic training networks ($\theta_{\mu}, \theta_{Q_1}$) and Actor/critic target networks ($\theta_{\mu}^-,\theta_{Q_1}^-$) with value estimation governed by
\begin{equation}\label{rt6}
y_t^{\text{DDPG}}=R_{t+1}+ \gamma^* Q(s_{t}^{i,j},a_t^{i,j},\theta_{Q_1}^-),
\end{equation}
where $\gamma^*$ is the discount factor.

To mitigate the inherent Q-value overestimation of DDPG, TD3 introduces dual Q-networks ($\theta_{Q_2},\theta_{Q_2}^-$) with conservative value estimation as
\begin{equation}\label{rt7}
y_t^{\text{TD3}}=R_{t+1}+ \gamma^* \min_{m=1,2} Q(s_{t}^{i,j},a_t^{i,j},\theta_{Q_m}^-).
\end{equation}

The loss function of the actor training network for the TD3 algorithm can be expressed as
\begin{equation}\label{rt8}
loss_{\text{TD3}}(\theta_{Q_m})=[y_t^{\text{TD3}}-Q(s_t^{i,j},a_t^{i,j},\theta_{Q_m})]^2,m=1,2.
\end{equation}

The loss function of the critic training network for the TD3 algorithm can be expressed as
\begin{equation}\label{rt88}
loss_{\text{TD3}}(\theta_{\mu})=Q(s_t^{i,j},a_t^{i,j},\theta_{Q_1}).
\end{equation}

The target network shares the same architecture as the training network, and the training network updates its own parameters to the target network after some time step. The expression for the update process can be shown as
\begin{equation}\label{rt9}
\begin{aligned}
\theta_{\mu}^- = \tau_1 \theta_{\mu} + (1- \tau_1) \theta_{\mu}^-, \\
\theta_{Q_1}^-= \tau_1 \theta_{Q_1} + (1- \tau_1) \theta_{Q_1}^-, \\
\theta_{Q_2}^-= \tau_1 \theta_{Q_2} + (1- \tau_1) \theta_{Q_2}^-, \\
\end{aligned}
\end{equation}
where $\tau_1 \in (0,1)$ is the soft update coefficient. It indicates that the target network retains the previous target network parameters with a weight of $1-\tau_1$, and updates the parameters of the training network to the target network with a weight of $\tau_1$.
% \IEEEpubidadjcol

\subsection{Twin-Dueling Multi-Agent Algorithm}
This paper presents a multi-agent algorithmic framework that integrates both TBA and DUA to collaboratively address the multi-dimensional optimization problem in FA-assisted MEC offloading systems. The proposed algorithm builds upon two key enablers, IBM-CCS channel estimation and game-theoretic power allocation, which together lay the foundation for accurate system awareness and resource efficiency. IBM-CCS ensures high-fidelity yet low-overhead channel state acquisition in FA scenarios, while game theory supports distributed and fair power control across users. These modules significantly reduce the uncertainty and dimensionality of the environment, thereby creating favorable conditions for effective agent training and decision making. To address the remaining challenges, we propose an IBM-CCS and game theory assisted HiTDMA. The adoption of a multi-agent framework overcomes the intrinsic limitations of single-agent systems, where isolated decision-making often fails to scale or converge in high-complexity scenarios.

In this framework, the original high-dimensional action space is decomposed into task-specific subspaces. Each agent operates within a significantly reduced decision domain, enhancing policy learnability and convergence. Crucially, the framework introduces an information-sharing mechanism where each agent can share important information, thus solving the convergence difficulty caused by partial observability in traditional multi-agent systems. This collaborative framework considers the interdependence of agent decisions in complex optimization scenarios, requiring each agent to dynamically adjust its strategy by considering environmental feedback and other agent behaviors. Agents in this system share key operational information with each other, enabling coordinated decision-making that considers both individual goals and group interactions.
\begin{algorithm}
    \caption{IBM-CCS and Game Theory-Assisted HiTDMA for Computation Offloading}\label{alg}
    \begin{algorithmic}[1]
        \STATE$\mathbf{Initialize:}$ Environment parameters.
        \STATE $\mathbf{Initialize:}$ Network parameters $\theta$ and $\theta^-$ of DUA, network parameters $\theta_{\mu}$, $\theta_{\mu}^-$, $\theta_{Q_1}$, $\theta_{Q_1}^-$, $\theta_{Q_2}$ and $\theta_{Q_2}^-$ of TBA.
        \FOR{Episode $n_e=1,2,\ldots,N_p$} 
            \STATE Initialize the position of FA ports.
            \FOR{Time slot $t = 1,2,\ldots,N_t$}
                \STATE Reset the environment. Utilize partial channel state to estimate the complete channel parameter vector $\tilde{\mathbf{g}}_n$ using the IBM-CCS method, and construct the user observation $s_t^u$ according to (\ref{channel_equation}).
                \STATE DUAs select actions $a_t^u =\left [ a_{t,1}^u,a_{t,2}^u,\ldots,a_{t,N}^u \right ]$.
                \STATE Get BS observation $s_t^b$ according to (\ref{eq:base_state}).
                \STATE TBAs select action $a_t^b$.
                \STATE Transform price factor $\lambda$ in action $a_t^b$ into power design $\mathbf{P}$ according to (\ref{power_iteration}), and get new action $\bar{a_t^b}=\left [ \text{Re}\{\mathbf{W}\}, \text{Im}\{\mathbf{W}\},\mathbf{P},\mathbf{Z} \right ]$.
                \STATE Executes action $\{a_{t,1}^u,a_{t,2}^u,\ldots,a_{t,N}^u\}$ and $\bar{a_t^b}$ get new state $s_{t+1}^u,s_{t+1}^b$ and recieve reward $r_t$ according to (\ref{eq:reward design}).
                \STATE Store the transitions $\left [s_t^u,a_t^u,r_t,s_{t+1}^u\right ]$ and $\left [s_t^b,a_t^b,r_t,s_{t+1}^b\right ]$ into the memory queues.
                \STATE The DUAs and TBAs calculate the objective function according to (\ref{rt5}) and (\ref{rt7}).
                \STATE The DUA calculates the loss function according to (\ref{rt66}). The TBA calculates the loss function according to (\ref{rt8}) and (\ref{rt88}). Calculate the gradient to update the training network parameters.
                \STATE The DUA updates the target network parameters of D3QN according to (\ref{eq10}) and the TBA update the target network parameters of TD3 according to (\ref{rt9}).
                \STATE Renew the environment information.
            \ENDFOR
        \ENDFOR
    \end{algorithmic}
\end{algorithm}

Specifically, the algorithm leverages game-theoretic modeling to simplify the high-dimensional user power allocation subproblem, while the IBM-CCS module provides accurate and structured channel feedback, enabling agents to perceive and adapt to the FA channel dynamics more effectively. The overall optimization is then performed via a hierarchical DRL combining DUA for discrete decision-making tasks and TBA for continuous control tasks. Each user utilizes one D3QN-based DUA responsible for learning the port selection strategy of the FA, while the BS utilizes three distinct TD3-based TBAs to handle beamforming, price factor selection, and MEC resource allocation strategy of the receiver, respectively. The algorithm operates in two phases, the training stage and the model inference stage. The detailed procedure of the proposed algorithm is summarized in Algorithm~\ref{alg}.
\subsubsection{Training Stage}
During the training stage, TBAs and DUAs initially exchange partial historical information to determine current environmental states. DUAs then select FA ports while TBAs implement beamforming for receivers, set price factors, and allocate MEC resources according to current strategies. To enhance exploration, DUAs probabilistically explore all actions while TBAs introduce adjustable noise in the decision processes. Throughout iterative training cycles, DUAs progressively refine their D3QN models through environmental feedback and reward, while TBAs correspondingly optimize TD3 models. This dual-update mechanism drives the system toward stable policy convergence through coordinated learning.
\subsubsection{Model Inference Stage}
In the model inference stage, agents cease strategy updates and instead utilize converged models to process real-time environmental observations. The trained models generate optimized actions directly from current states, enabling efficient computational offloading. This operational shift ensures rapid decision-making while maintaining the coordinated optimization capabilities developed during training, effectively balancing exploration benefits from the learning phase with execution efficiency requirements during deployment.

\section{Numerical Results}
This section presents a comprehensive evaluation of the proposed algorithm through simulation experiments, focusing on both channel estimation accuracy and computation offloading optimization performance in FA-assisted MEC systems. The experimental configuration follows the parameter settings outlined in Table \ref{t1}, establishing a comprehensive testing environment for algorithm verification. The effectiveness of channel estimation performance is assessed using two widely adopted evaluation metrics~\cite{sara2019image}. The first metric is Peak Signal to Noise Ratio (PSNR), which reflects the fidelity of the reconstructed channel relative to the original channel state. The second metric is Structural Similarity (SSIM), which measures the ability of the reconstruction to preserve spatial structure and correlation patterns within the estimated channel matrices. For DRL-enabled optimization algorithm assessment, offloading effectiveness is evaluated in terms of total delay, which captures the joint effect of task transmission and edge processing. The proposed algorithm is compared against multiple benchmark schemes to assess its advantage in learning efficiency, coordination capability, and adaptability to the FA channel environment. Benchmark algorithms and performance metrics undergo comparative analysis under standardized evaluation conditions.
\subsection{Simulation Setup}
\begin{table}[!htbp]
\centering
\caption{Simulation Parameters\vspace{-0.2cm}} % 添加标题
\begin{tabular}{|l|c|}%l:left c:center r:right 
\hline 
\textbf{Parameter} & \textbf{Value} \\
% \hline 
% Number of users $N$ & 3  \\
% \hline 
% Number of FA ports $M$ & 32 \\
\hline 
Height of FA BS station $h$ (m) & 15 \\
\hline 
Wavelength $l$ (m) & 0.1 \\
\hline 
Bandwidth $B$ (GHz) & 1 \\
\hline 
Noise power $\sigma^2$ (dBm) & -84 \\
\hline 
Trade-off parameter $\eta$ & $10^{-2}$ \\
\hline 
Regularization parameter $\gamma$ & $10^{-4}$ \\
% \hline 
% Compressive sensing ratio $C$ & 0.1 \\
\hline
Sampling ratio $r$ & 0.1\\
\hline
Block size $B_k$ & 32\\
\hline
Kernel size $f$ & 3\\
\hline
Number of feature map $d$ & 64\\
\hline
Number of layers in deep reconstruction network $n$ & 5\\
\hline 
Maximum transmit power of user $p_{max}$ (dBm)& 17 \\
\hline 
Maximum computation capacity of user $f_{max}^l$ (MHz)& 1 \\
\hline 
Maximum computation capacity of MEC server $F^{max}$ (MHz) & 100 \\
\hline
Positive constant $\varphi$ & 1\\
\hline
Positive constant $\nu$ & 5\\
\hline
\end{tabular}
\label{t1}
\end{table}
% The trade-off parameter between estimating the target well and limiting the information $\eta$ is set to $10^{-2}$ in the importance generator training process. The regularization parameter $\gamma$ is set to $10^{-4}$, and the compression sensing ratio $C$ is fixed at $0.1$ in sensing and decoder network training. The channel measurement data collected by the USRP device forms a training data set, and its specific network architecture parameters are as follows: block size $B_k =32$, convolutional kernel spatial dimension $f = 3$, the number of feature maps in depth reconstruction networks employ $d = 64$, and the number of nonlinear mapping cascaded layers is $n=5$. The MEC simulation environment implements three FAs at FA-BS, each supporting 32 ports. 
% The channel measurement data collected by the USRP device forms a training data set in CS network training, and its specific network architecture parameters are outlined in Table \ref{t1}. 
The neural architecture implementation specifies identical three-layer configurations for both D3QN and TD3 frameworks, with successive fully connected layers containing 64, 128, and 64 neurons, respectively. The D3QN component employs Rectified Linear Unit activation functions, while the TD3 module utilizes Sigmod activation patterns, both optimized through Adam adaptive gradient descent with a learning rate fixed at 0.0001. Training protocols implement gradient clipping with a threshold of 0.25 and a discount factor of 0.8 to ensure numerical stability during backpropagation. Exploration parameters undergo linear decay from the initial value 1 to the final value 0.02 across 800 training epochs, maintaining constant exploration thereafter. The experiments are implemented on an NVIDIA GTX 1660 GPU.

\subsection{Benchmark schemes}
To conduct a comprehensive performance analysis, the proposed computation offloading scheme is evaluated against several benchmark schemes under identical simulation settings. The comparison schemes are described below.
\begin{itemize}
    \item [*]
    \textbf{CCS}. This scheme maintains identical neural network structure as the IBM-CCS framework while deliberately excluding the importance generator module. 
    \item [*]
    \textbf{S-BAR}~\cite{zhangSuccessiveBayesianReconstructor2023}. This scheme employs a successive Bayesian reconstruction approach that leverags Gaussian process regression with experiential kernels for channel estimation. It serves as a classical learning-based baseline for compressive channel recovery.
    \item [*]
    \textbf{Fixed position antenna scheme (FPA)}. This scheme replaces FA with FPA at the BS and removes DUAs while retaining TBAs for optimization of beamforming, power design, and MEC resource allocation. The scheme operates under IBM-CCS-based channel estimation to ensure consistency in channel input across all comparisons.
    \item [*]
    \textbf{Fixed power scheme (FP)}. This scheme employs deterministic user transmission power while maintaining APV optimization via DUAs and optimization of beamforming and MEC allocation through TBAs. The optimization processes are executed based on the channel state estimated by the IBM-CCS.
    \item [*]
    \textbf{Zero forcing scheme (ZF)}. This scheme leverages classical zero-forcing techniques for beamforming matrix generation while preserving  APV optimization, power control, and MEC allocation. The entire optimization is carried out under IBM-CCS-based estimated channel conditions.
    \item [*]
    \textbf{Multi-agent deep deterministic policy gradient (MADDPG)}~\cite{lowe2017multi}. In this scheme, the DUAs and TBAs in the proposed framework are replaced by DDPG-based agents. The optimization is carried out under IBM-CCS-based estimated channel conditions.
\end{itemize}

\subsection{Performance Evaluation}

% \begin{figure}
%     \centering
%     % \subfigcapskip=-5pt
%     % \subfigbottomskip=-2pt
% 	\subfigure[Convergence performance of PSNR for each training episode.]{
% 		\begin{minipage}[t]{1\linewidth}
% 			\includegraphics[width=3.3in]{figure/PSNR_Training.eps}
% 		\end{minipage}
% 	}%
 
% 	\subfigure[Convergence performance of SSIM for each training episode.]{
% 		\begin{minipage}[t]{1\linewidth}
% 			\includegraphics[width=3.3in]{figure/SSIM_Training.eps} 
% 		\end{minipage}  
% 	}%
% \caption{Convergence performance for each training episode.}
% \label{f8}
% \end{figure}

\begin{figure}[htbp]
    \centering
    \begin{subfigure}{0.48\linewidth}
      \includegraphics[width=\linewidth]{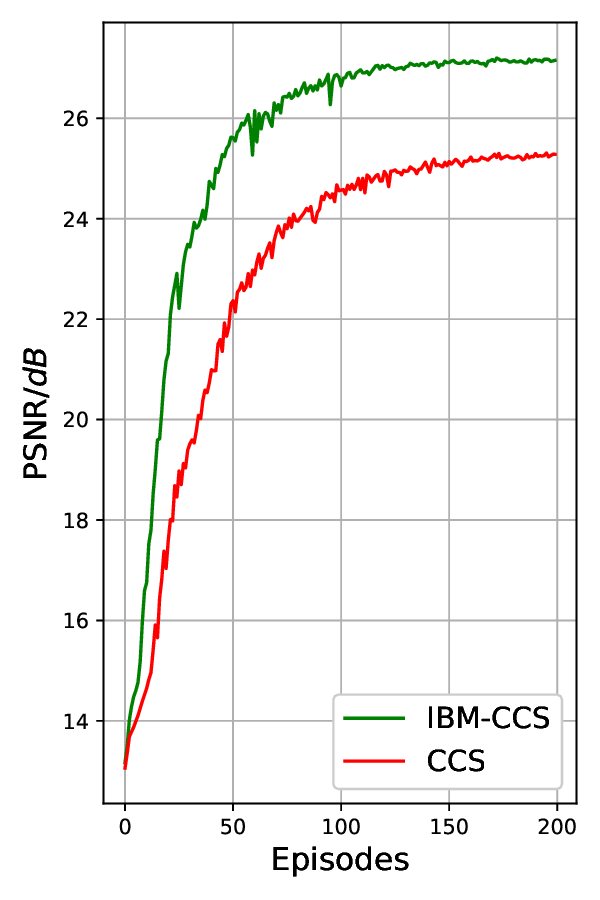}
      \caption{Convergence performance of PSNR for each training episode.}
      \label{f81}
    \end{subfigure}
    \hfill
    \begin{subfigure}{0.48\linewidth}
      \includegraphics[width=\linewidth]{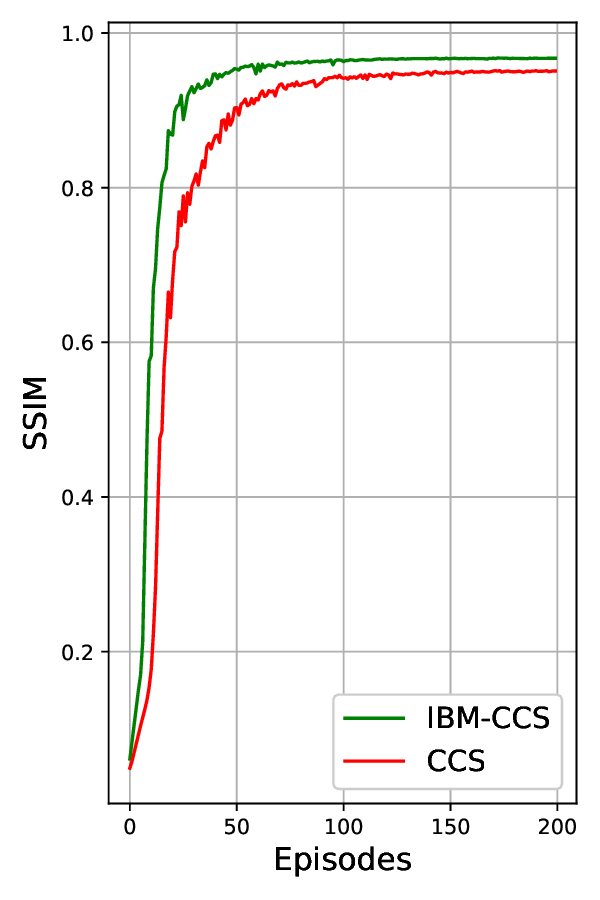}
      \caption{Convergence performance of SSIM for each training episode.}
      \label{f82}
    \end{subfigure}
    \caption{Convergence performance for each training episode.}
    \label{fig:overall}
  \end{figure}

Fig.~\ref{fig:overall}~(a) exhibits the comparative analysis of PSNR convergence trajectories between the proposed IBM-CCS channel estimation method and the approach without IBM, revealing significant performance advantages in both convergence dynamics and steady-state precision. The proposed method demonstrates accelerated convergence characteristics, reaching a stable PSNR plateau within approximately 30\% fewer iterations compared to the benchmark, which exhibits prolonged oscillatory behavior before stabilization. Notably, the proposed scheme achieves superior final convergence quality, maintaining a sustained PSNR advantage of 1.2-1.8 dB in steady-state operation. This performance gap emerges early in training, then progressively widening as training progresses. The method also exhibits enhanced stability. These advantages collectively confirm the enhanced learning efficiency and noise resilience of the proposed channel estimation framework, which is especially beneficial in FA-enabled environments where rapid channel variation demands fast and stable estimation. Such characteristics are particularly valuable for real-time implementations requiring both high convergence speed and reliable performance under dynamic FA configurations.

Fig.~\ref{fig:overall}~(b) exhibits the analysis of SSIM convergence patterns under FA configuration (length=$0.5l$, 256 ports), demonstrating the superior structural information preservation capabilities and accelerated learning dynamics of the proposed IBM-CCS channel estimation scheme compared to the CCS scheme. The proposed shceme achieves rapid early-stage convergence, reaching 90\% of its final SSIM value within 40 iterations which is 25\% faster than the benchmark. Notably, the method establishes a persistent performance advantage throughout training, maintaining a 3.8-8.2\% absolute SSIM lead across all phases. The enhanced stability is evidenced by reduced oscillation magnitude. These advantages collectively validate the method's enhanced capability in maintaining structural coherence during channel estimation, a critical requirement in FA scenarios where spatial diversity and antenna mobility necessitate robust signal structure preservation.

% \begin{figure}[htbp]
%     \centering
%     \includegraphics[width=0.48\textwidth]{figure/PSNR_Training.eps} % 调整宽度
%     \caption{.}
%     \label{fig:psnr_train}
% \end{figure}
% \begin{figure}[htbp]
%     \centering
%     \includegraphics[width=0.48\textwidth]{figure/SSIM_Training.eps} % 调整宽度
%     \caption{.}
%     \label{fig:ssim_train}
% \end{figure}

\begin{figure}[htbp]
    \centering
    \includegraphics[width=0.48\textwidth]{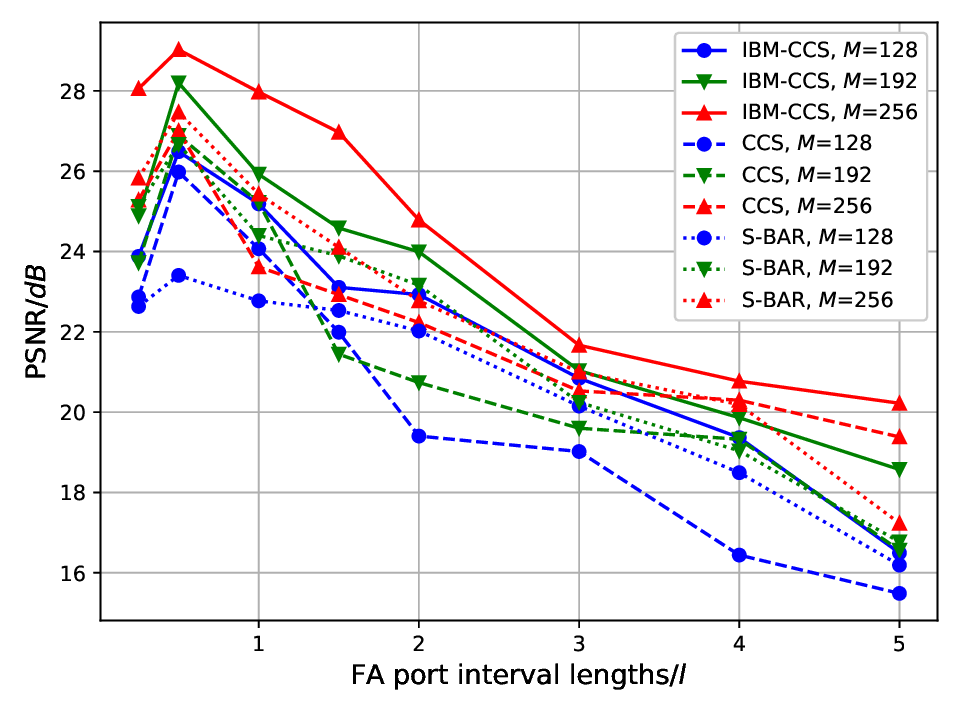} % 调整宽度
    \caption{PSNR performance under different FA port interval lengths.}
    \label{fig:psnr}
\end{figure}
The analysis of PSNR performance across varying FA port interval lengths in Fig.~\ref{fig:psnr} reveals distinct advantages of the proposed channel estimation method over the benchmark approach. For all port numbers (128, 192, 256), the IBM-CCS method consistently achieves higher PSNR values compared to both CCS and S-BAR, particularly in critical low-interval regimes below $0.5l$ where mutual coupling effects dominate. When the port spacing decreases to $0.25l$, performance degradation is observed in all approaches, stemming from fundamental electromagnetic constraints that closer spacing exacerbates mutual coupling effects and spatial aliasing. In this case, IBM-CCS achieves a PSNR of 28.06 dB, compared to 25.28 dB for CCS and 25.83 dB for S-BAR. This highlights the strength of IBM-CCS in capturing key spatial features through its learned importance weighting mechanism. As the antenna spacing increases, PSNR values decline across all methods due to reduced spatial coherence. However, IBM-CCS maintains a performance margin over others. S-BAR performs well but is limited by its kernel-based regression approach, which lacks adaptive feature prioritization. CCS, while sharing the base architecture of IBM-CCS, suffers from its omission of the importance generator, resulting in suboptimal information retention during compression. Overall, IBM-CCS demonstrates strong generalization and robustness to varying FA conditions, making it a superior choice for channel estimation in dynamic FA environments. Integrating it into the offload optimization framework has the potential to improve downstream decision quality and system efficiency.

\begin{figure}[htbp]
    \centering
    \includegraphics[width=0.48\textwidth]{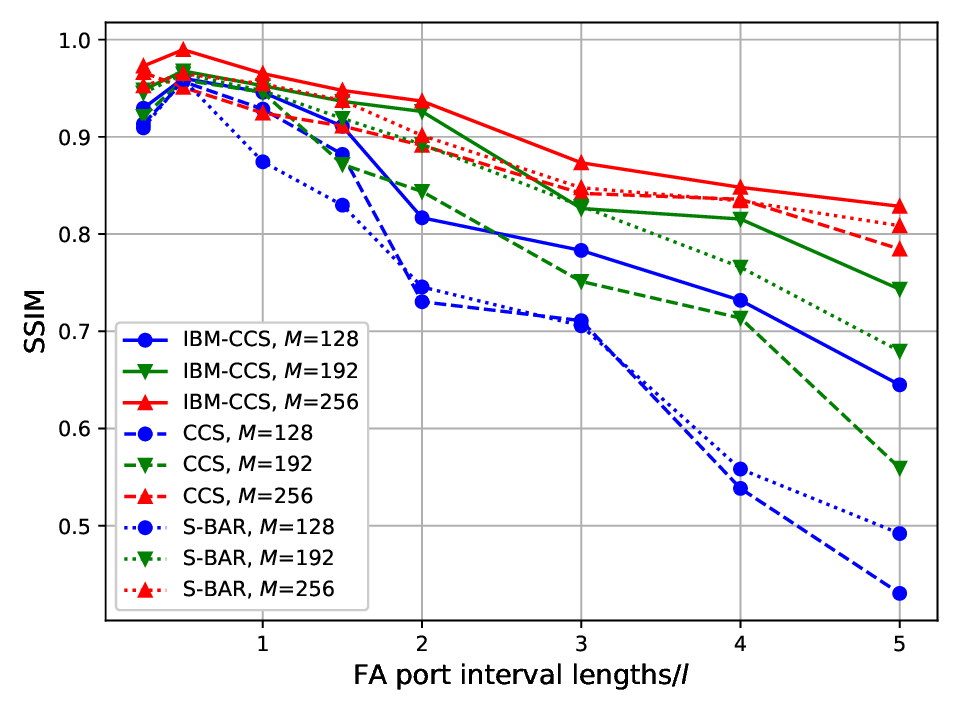} % 调整宽度
    \caption{SSIM performance under different FA port interval lengths.}
    \label{fig:ssim}
\end{figure}
The comparative analysis of SSIM performance across varying FA port interval lengths in Fig.~\ref{fig:ssim} demonstrates significant superiority of the proposed IBM-CCS channel estimation method over other baseline approaches in preserving structural information integrity. For all port configurations, the proposed IBM-CCS method maintains consistently higher SSIM values throughout the FA port interval length spectrum. 
% At 256 ports, which represent high-density scenarios, the proposed technique achieves exceptional performance retention, maintaining SSIM above 0.8285 at maximum length ($5l$) compared to the baseline method's 0.7846, while establishing a substantial 0.63\% absolute advantage (0.9728 vs 0.9662) at minimal antenna dimensions ($0.25 l$). 
Notably, similar to the above PSNR performance analysis, when the port spacing decreases to $0.25l$, where mutual coupling effects and spatial correlation are strongest, IBM-CCS delivers clear advantages. For instance, with 256 ports and 0.5 spacing, IBM-CCS attains an SSIM of 98.9\%, significantly higher than CCS at 95.0\% and S-BAR at 96.4\%. This highlights the effectiveness of the importance generator in selectively preserving critical spatial features during compressive sensing. As the antenna spacing increases, SSIM values generally decline across all methods due to weakened spatial coherence. However, IBM-CCS maintains superior structural preservation even at larger interval spacings. Overall, IBM-CCS demonstrates robust generalization across FA configurations, offering high structural fidelity even under challenging conditions. Likewise, the reliability in channel representation directly benefits downstream optimization tasks in FA-assisted MEC systems.

\begin{figure}[htbp]
    \centering
    \includegraphics[width=0.48\textwidth]{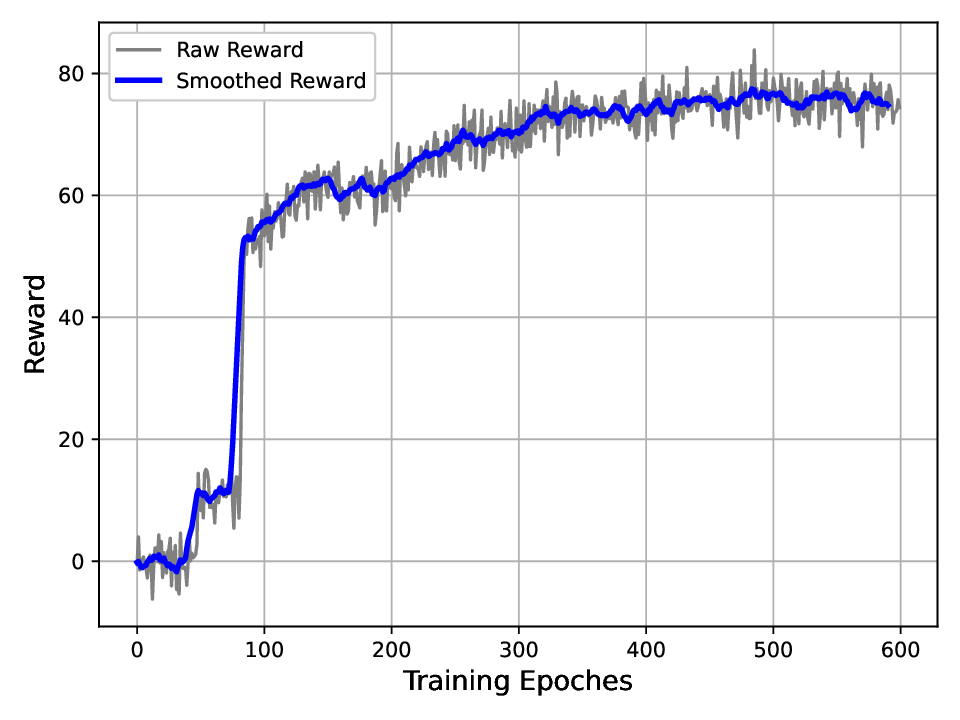} % 调整宽度
    \caption{Cumulative reward for each training episodes.}
    \label{fig:iter}
\end{figure}

% Fig. \ref{fig:iter} exhibits the relationship between the training epochs of the multi-agent model and the cumulative reward, which shows the convergence of the game theory-assisted MADDPG algorithm. The figure reveals that before the 300 iterations, the model is learning by exploring the environment, which makes its reward value small and unstable. After 250 iterations, the obtained reward is high and gradually becomes stable, which means that when the environment changes, the model has a stable strategy to obtain a higher reward. 

Fig.~\ref{fig:iter} exhibits the reward convergence of the proposed DRL-based optimization method, which incorporates a FAS, demonstrating a clear upward trend that reflects effective learning. In early training, rewards fluctuated within 30 iterations with high variance. As training proceeds, rewards steadily increase, surpassing 60 and eventually stabilizing around 88–90, indicating successful policy optimization. A sharp rise between iterations 40 and 70 marks a key learning phase. The reduced variance and high reward density in later stages confirm robust convergence. Compared to conventional methods, the proposed approach benefits from the synergy between FA and DRL, which enhances adaptability and spatial awareness, enabling more efficient exploration and higher-quality solutions. The reward trajectory and final performance underscore the effectiveness of the proposed method.

% \begin{figure}[htbp]
%     \centering
%     \includegraphics[width=0.48\textwidth]{figure/img_iter0.eps} % 调整宽度
%     \caption{.}
%     \label{fig:iter0}
% \end{figure}

\begin{figure}[htbp]
    \centering
    \includegraphics[width=0.48\textwidth]{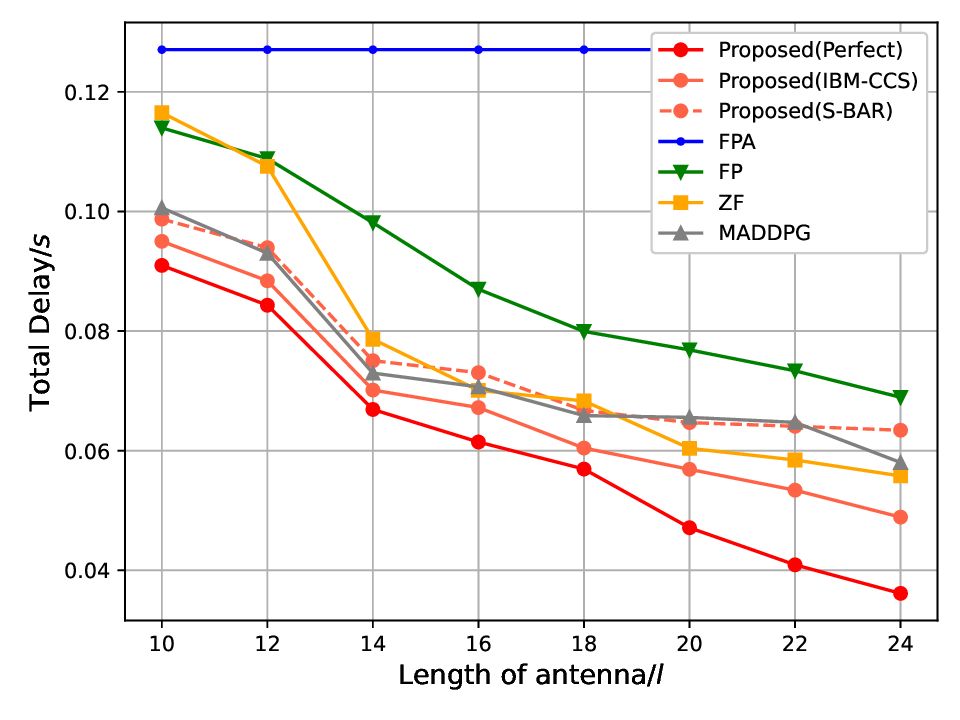} % 调整宽度
    \caption{System delay under the FA length with $N = 3$, and $M = 32$.}
    \label{fig:length}
\end{figure}

% Fig. \ref{fig:length} illustrates the performance of the proposed scheme in the FA-aided MEC offloading scenario, demonstrating its effectiveness compared with the benchmark schemes under various antenna lengths. It reveals that as the length of FA increases and the number of FA ports remains unchanged, the system delay of all schemes except the FPA scheme shows a downward trend. Compared with the FPA scheme, the other FA-aided schemes have lower system delay overall, mainly due to their effective improvement of the transmission rate by employing the extraordinary multiplexing gain of FA. Moreover, the comparison with the proposed scheme and ZF scheme, FP scheme reveals the beneficial impact on the optimization of beamforming and user power design respectively. 

Fig.~\ref{fig:length} illustrates the performance evaluation of the FA-assisted MEC offloading framework under varying antenna lengths. The proposed optimization algorithm is evaluated under both ideal conditions with perfect channel state information and realistic conditions using IBM-CCS and S-BAR-based channel estimation. All benchmark methods, including FPA, FP, ZF, and MADDPG are implemented under IBM-CCS-based estimated channels. Results indicate that the proposed algorithm consistently outperforms all benchmark methods and is closest to the scheme under the perfect channel. As the antenna length increases from $10l$ to $24l$, the total delay achieved by the proposed method exhibits a clear and steady downward trend, highlighting the capability to leverage the spatial diversity and dynamic reconfigurability of FA efficiently. Furthermore, the proposed method maintains a significant performance advantage over the S-BAR-based baseline across all antenna lengths, suggesting that IBM-CCS provides sufficiently accurate channel reconstruction to support efficient offloading decisions. In contrast, the FPA yields a constant and significantly higher delay, underscoring the performance loss incurred by replacing reconfigurable FA with fixed antennas. While FP and ZF also exhibit a downward trend, their delay reductions are neither as significant nor as consistent as the proposed method, revealing the beneficial impact of the proposed method on the optimization of user power design and beamforming respectively. Furthermore, compared to MADDPG, the proposed method provides lower delays across the entire range, particularly at larger FA lengths where spatial reconfigurability becomes more prominent, demonstrating the robustness of the proposed optimization strategy. In summary, the results validate that the proposed algorithm effectively enhances MEC offloading efficiency by integrating IBM-CCS-based channel estimation with adaptive FA port selection, beamforming design, user power control, and MEC resource allocation in a unified and robust learning framework.
% even under imperfect channel conditions, where the proposed method operates using estimated CSI derived from the novel estimation technique introduced in this work, the delay performance still surpasses all benchmark schemes under perfect CSI. The delay under estimation conditions improves steadily with FA length. This not only demonstrates the robustness of the proposed optimization strategy to channel uncertainties but also affirms the effectiveness of the developed channel estimation method in enabling near-optimal decision-making under realistic conditions. Overall, these results clearly validate the superiority of the proposed method in enhancing MEC offloading efficiency by dynamically adapting the FA port selection, beamforming, user transmit power, and MEC offloading resource allocation.

\begin{figure}[htbp]
    \centering
    \includegraphics[width=0.48\textwidth]{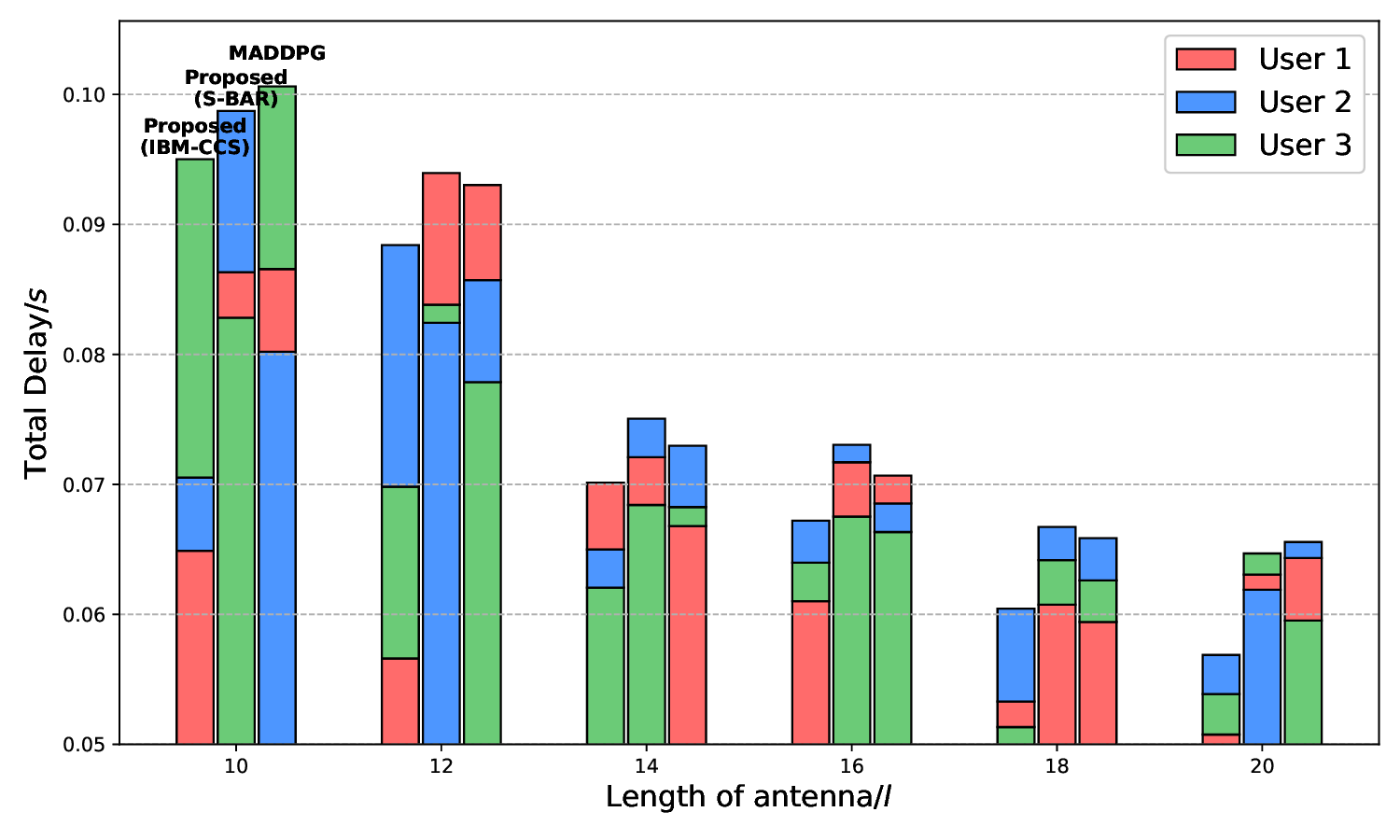} % 调整宽度
    \caption{Delay of all users under the FA length with $N = 3$, and $M = 32$.}
    \label{fig:all_user}
\end{figure}

Fig.~\ref{fig:all_user} illustrates the offloading delay performance of all users under varying antenna lengths, comparing the proposed algorithm under IBM-CCS and S-BAR CSI estimation conditions, as well as the MADDPG baseline under IBM-CCS. It is evident that the proposed algorithm under IBM-CCS estimation consistently achieves the lowest and most stable delay across all antenna length configurations. Notably, under the IBM-CCS condition, the proposed method not only minimizes the maximum offloading delay among users, but also ensures uniformly low delays across all users. Compared to the MADDPG, the proposed approach not only achieves lower delays but also exhibits less variance among users, indicating better fairness and load balancing in the offloading process. When contrasted with the performance under S-BAR estimation, the IBM-CCS-based results show a clear advantage, particularly in shorter antenna length configurations where the delay reduction is more significant. These results highlight the dual superiority of the proposed algorithm, which is that the IBM-CCS channel estimation framework provides accurate and informative channel features for optimization, while the tailored offloading strategy effectively leverages these features to achieve globally balanced and efficient computation offloading.

\begin{figure}[htbp]
    \centering
    \includegraphics[width=0.48\textwidth]{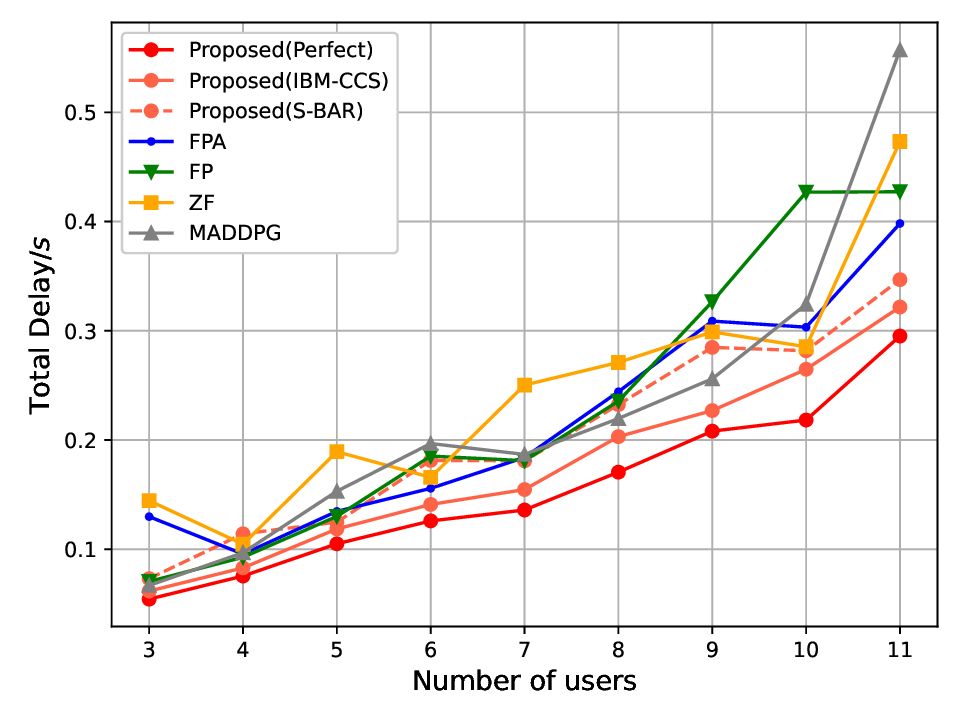} % 调整宽度
    \caption{System delay under the number of users with $W = 16$, and $M = 32$.}
    \label{fig:user}
\end{figure}

% Fig. \ref{fig:user} illustrates the effect of the number of users on the total delay in different schemes. We can observe that as the number of users increases, the proposed scheme is superior to other benchmark schemes, and the delay increases slowly so that it can cope with scenarios with more users. Overall, the performance of FA-aided schemes is much better than the FPA scheme. Optimizing beamforming and user power design also has a beneficial effect on reducing system delay.

Fig.~\ref{fig:user} illustrates the effect of the number of users on the total delay in different schemes, which reveals the substantial advantage of the proposed optimization scheme in minimizing total delay for MEC offloading, especially in multi-user scenarios. As the number of users increases, the delay associated with the proposed method gradually increases, reflecting the growing computational burden and inter-user interference. Nonetheless, the proposed method consistently outperforms all benchmark schemes across all user counts except the scheme under perfect CSI, confirming its scalability and efficient resource coordination in dense multi-user environments. The performance of the proposed method is consistently lower than that observed in S-BAR-based benchmarks, indicating that the proposed IBM-CCS estimation delivers channel fidelity sufficient for effective offloading decisions, even under dynamic user conditions. In contrast, the FPA scheme shows a steep increase in delay, and the delay is higher than that of all FA-equipped schemes, which highlights the high efficiency of FA in adapting to user density. Similarly, both FP and ZF experience notable degradation in performance, and there is a considerable gap with the proposed scheme, reflecting their limited adaptability in high-load scenarios without comprehensive joint optimization. Compared to MADDPG, the proposed algorithm demonstrates improved robustness and lower delay across all user counts. This advantage becomes more evident as the system scales, showcasing the superiority of the HiTDMA. Overall, this analysis underscores the superior performance, scalability, and resilience of the proposed method in complex and dynamic MEC environments.
% Furthermore, even with the imperfect CSI estimated by the proposed channel estimation approach, the "Proposed-nonPerfect" results continue to outperform most benchmarks with perfect CSI in most scenarios. The relatively smooth delay growth of the "Proposed-nonPerfect" scheme across increasing user numbers highlights both the robustness of the proposed optimization scheme and the reliability of the channel estimation technique. Overall, this analysis underscores the superior performance, scalability, and resilience of the proposed method in complex and dynamic MEC environments.

\section{Conclusion}
This paper presents an integrated FA-assisted MEC offloading framework that combines a hierarchical MADRL-based optimization strategy with a novel IBM-CCS channel estimation method. Comprehensive evaluations demonstrate that the proposed system consistently achieves superior performance and robustness across diverse deployment scenarios. In particular, IBM-CCS outperforms conventional approaches in terms of PSNR and SSIM under different antenna configurations, ensuring reliable channel reconstruction even under low FA-port densities. These high-fidelity estimations enable the DRL optimization to achieve offloading delays close to those under perfect CSI, thereby confirming the practicality of the framework in dynamic and resource-constrained environments.
Moreover, the proposed HiTDMA demonstrates strong scalability with increasing antenna lengths and user numbers, maintaining lower delay than all benchmarks. This demonstrates its effectiveness in exploiting FA reconfigurability and spatial diversity. Overall, the close synergy between accurate channel estimation and adaptive multi-agent optimization provides a robust and efficient offloading solution, making the framework well-suited for MEC networks.

\bibliographystyle{IEEEtran}
\bibliography{IEEEabrv, reference/reference}

% \newpage

% \section{Biography Section}
% If you have an EPS/PDF photo (graphicx package needed), extra braces are
%  needed around the contents of the optional argument to biography to prevent
%  the LaTeX parser from getting confused when it sees the complicated
%  $\backslash${\tt{includegraphics}} command within an optional argument. (You can create
%  your own custom macro containing the $\backslash${\tt{includegraphics}} command to make things
%  simpler here.)
 
% \vspace{11pt}

% \bf{If you include a photo:}\vspace{-33pt}
% \begin{IEEEbiography}[{\includegraphics[width=1in,height=1.25in,clip,keepaspectratio]{fig1}}]{Michael Shell}
% Use $\backslash${\tt{begin\{IEEEbiography\}}} and then for the 1st argument use $\backslash${\tt{includegraphics}} to declare and link the author photo.
% Use the author name as the 3rd argument followed by the biography text.
% \end{IEEEbiography}

% \vspace{11pt}

% \bf{If you will not include a photo:}\vspace{-33pt}
% \begin{IEEEbiographynophoto}{John Doe}
% Use $\backslash${\tt{begin\{IEEEbiographynophoto\}}} and the author name as the argument followed by the biography text.
% \end{IEEEbiographynophoto}

\vfill

\end{document}